\documentclass[a4paper,fleqn]{cas-dc}

\usepackage[authoryear]{natbib}

\def\tsc#1{\csdef{#1}{\textsc{\lowercase{#1}}\xspace}}
\tsc{WGM}
\tsc{QE}
\tsc{EP}
\tsc{PMS}
\tsc{BEC}
\tsc{DE}

\usepackage{colortbl}
\usepackage{xcolor}
\usepackage[ruled, vlined]{algorithm2e}
\usepackage{algpseudocode}
\usepackage{bm}
\usepackage{amsmath}
\DeclareMathOperator*{\argmax}{arg\,max}

\def\d{{\mbox{d}}}
\def\E{{\mathbb E}}
\usepackage{mathtools}

\begin{document}
\let\WriteBookmarks\relax
\def\floatpagepagefraction{1}
\def\textpagefraction{.001}

\shorttitle{Hybrid Model Assisted Reinforcement Learning for Cell Therapy Manufacturing}

\shortauthors{Zheng et~al.}

\title [mode = title]{Opportunities of Hybrid Model-based Reinforcement Learning for Cell Therapy Manufacturing Process Control}




%
\author[1]{Hua Zheng}






\address[1]{Department of Mechanical and Industrial Engineering, Northeastern University, Boston, MA 02115, USA}

\author[1]{Wei Xie}[orcid=0000-0001-9563-4927]
\ead{w.xie@northeastern.edu}
\cormark[1]
\author[1]{Keqi Wang}

\author[3]{Zheng Li}
\address[3]{Genentech, Inc., South San Francisco, CA, USA}


\cortext[cor1]{Corresponding author}



\begin{abstract}
Driven by the key challenges of cell therapy manufacturing, including high complexity, high uncertainty, and very limited process observations, we propose a hybrid model-based reinforcement learning (RL) to efficiently guide process control. We first create a probabilistic knowledge graph (KG) hybrid model characterizing the risk- and science-based understanding of biomanufacturing process mechanisms and quantifying inherent stochasticity, e.g., batch-to-batch variation. It can capture the key features, including nonlinear reactions, nonstationary dynamics, and partially observed state. This hybrid model can leverage existing mechanistic models and facilitate learning from heterogeneous process data. A computational sampling approach is used to generate posterior samples quantifying model uncertainty. Then, we introduce hybrid model-based Bayesian RL, accounting for both inherent stochasticity and model uncertainty, to guide optimal, robust, and interpretable dynamic decision making. Cell therapy manufacturing examples are used to empirically demonstrate that the proposed framework can outperform the classical deterministic mechanistic model assisted process optimization.
\end{abstract}


\maketitle

\section{Introduction}
\label{sec:introduction}

\begin{sloppy}


The past decade has seen unparalleled growth in the field of cell therapies that are used to 
treat and prevent 
diseases, such as cancers, cardiovascular and hematologic diseases,
through the pharmacological, immunological, or metabolic actions of cells or tissues \citep{wang2021cell,hanna2016advanced}.
Kymriah (tisagenlecleucel) made history in August 2017 when it became the world's first approved chimeric antigen receptor (CAR) T-cell therapy. 
The projected worth of the cell therapy market is \$8.21 billion in 2025 \citep{fiorenza2020value}.

\textit{Cell therapy manufacturing often faces critical challenges, including high complexity, high variability, and very limited number of process observations.} 
The productivity and functional identity of cell products are sensitive to cell culture conditions. Improper cultivation can not only hinder yield, but can result in heterogeneously differentiated cell populations or contamination, which could give rise to potential tumor or teratoma formulation in cell graft recipients \citep{dressel2011effects}.
In addition, since cell therapies become more and more personalized, there are often very limited data for process development and control.
The seed cells can be extracted and isolated from individual patients and donors, which leads to high variability. 

For biochemical processes, the laws of biophysical chemistry often allow us to construct mechanistic models. 
However, existing mechanistic models often ignore the impact from various sources of biomanufacturing process inherent stochasticity. For example, although batch-to-batch variation and bioprocess noise are often dominant sources of process variation \citep{mockus2015batch}, biochemical kinetics literature rarely incorporates them into the ordinary/partial  differential equation (ODE/PDE) based mechanistic models.
In a study on the microbial cell-to-cell phenotypic diversity, \cite{vasdekis2015origins} identified the intracellular production fluctuations as one of the major sources of the bioprocessing noise. Raw material variability is another 
critical source of uncertainty impacting cell cultures \citep{dickens2018biopharmaceutical}. 


There are two types of uncertainty,
including (1) 
\textit{inherent stochasticity}, such as uncertainties from raw materials, critical process parameters (CPPs), and other uncontrolled variables (such as contamination); and (2)
\textit{model uncertainty} incurred because the bioprocess model is an approximation of the real process. 
The process inherent stochasticity can be controlled by improving the identification and specification of CPPs. 
The model uncertainty can be reduced by collecting more informative process observations. Given very limited process data, model uncertainty can be large.
\textit{Thus, correctly quantifying all sources of uncertainty can facilitate mechanism learning, guide risk reduction, and support robust 
biomanufacturing process optimization.}

Various process analytical technologies (PATs) and control methodologies 
have been proposed to guide biomanufacturing process decision making and variation control; see the review in \cite{steinwandter2019data,yoo2021reinforcement}. 
The classical feed-forward and feedback control strategies are often derived based on deterministic first-principle models \citep{hong2018challenges,kee2009selective,18cdc_tutorial_control_biology}. They typically overlook
bioprocess stochastic uncertainty and model uncertainty. Markov decision processes (MDP) and reinforcement learning (RL) based process optimization approaches draw increasing attentions in the recent years; such as \cite{spielberg2017deep,spielberg2020deep,martagan2018performance,liu2013modelling}. As pointed out in the recent review \citep{yoo2021reinforcement},
 existing biomanufacturing process optimization approaches tend to have several key limitations. First, mechanistic models, in which conservation equations, kinetics, thermodynamics, and transport phenomena are modeled by ordinary or partial differential equations (ODEs/PDEs), are usually deterministic, while real bioprocess operations are stochastic. 
Second, existing MDP and RL approaches often build on machine learning methodologies without incorporating the prior knowledge on bioprocessing mechanisms; for example black-box neural networks are used to approximate the value functions.
This limits their performance, as well as interpretability and 
sample efficiency.
Third, there is still lack of biomanufacturing process control strategies accounting for both inherent stochasticity and model uncertainty. This impacts mechanism learning and optimization robustness \citep{yoo2021reinforcement}.

Driven by the critical challenges faced by cell therapy manufacturing, we propose a data-driven stochastic optimization framework named ``hybrid-RL'' building on bioprocess probabilistic knowledge graph (KG) hybrid model in conjunction with model-based RL to efficiently guide optimal and robust process control. The key contributions of this paper are summarized as follows.
\begin{itemize}
    \item We first create a KG network hybrid model characterizing the risk- and science-based understanding of biomanufacturing process mechanisms and spatial-temporal causal interdependencies between CPPs and critical quality attributes (CQAs). It quantifies cell therapy manufacturing inherent stochasticity, including batch-to-batch variation and bioprocess noise. This hybrid model can leverage information from existing mechanistic models and further facilitate learning from process data. Comparing to existing approaches \citep{zheng2021policy}, the proposed knowledge graph hybrid model can capture the key features of bioprocesses, including nonlinear reactions, nonstationary dynamics, and partially observed state. 
    
    \item Given very limited and heterogeneous process observations, Bayesian inference is used to derive a posterior distribution of the process hybrid model. We use a computational sampling approach to generate posterior samples quantifying model uncertainty. In addition, our approach can provide long-term probabilistic prediction with prediction risk accounting for both inherent stochasticity and model uncertainty.
    
    \item We then introduce a hybrid model-based Bayesian RL framework (call ``hybrid-RL") to efficiently guide optimal, robust, and interpretable dynamic decision making and process control, 
  which can overcome the key challenges of cell therapy manufacturing. 
  The proposed model-based RL scheme on the Bayesian KG can provide an insightful prediction on how the effects of decision inputs 
  propagate through bioprocess mechanism pathways and impact on the output trajectory dynamics and variation. 
  
   \item The proposed KG hybrid model-based RL framework demonstrates promising performance for cell therapy manufacturing process optimization. 
   Our empirical studies show that this framework has a great potential to outperform the classical 
   mechanistic model assisted process optimization approach.
\end{itemize}


The remainder of the paper is organized as follows. We give a brief review of the biomanufacturing process modeling and optimization approaches in Section~\ref{sec:literatureReview}. 
We provide the problem description and summarize the proposed framework in Section~\ref{sec: problem description}. 
Then, we present a probabilistic KG hybrid model and use the approximate Bayesian computation (ABC) approach to 
generate posterior samples quantifying model uncertainty in Section~\ref{sec:hybridModeling}, which can facilitate bioprocess mechanisms learning. We further present the hybrid model based Bayesian RL with detailed algorithms in Section~\ref{sec:RL}. We conduct the empirical study of the erythroblast cell culture optimization 
in Section~\ref{sec:caseStudy} and conclude the paper in Section~\ref{sec:conclusion}. 

\end{sloppy}

\section{Literature Review}
\label{sec:literatureReview}

\begin{sloppypar}
\textbf{Bioprocess Modeling and Analysis:}
In classic biomanufacturing literature, the bioprocess dynamics is usually modeled by mechanistic models in forms of ordinary/partial differential equations (ODEs/PDEs) \citep{mandenius2013measurement}. 
Because they are capable of representing the complex biochemistry of cells in a more complete way 
\citep{almquist2014kinetic}, mechanistic models become popular and have been widely used in process optimization, design, control, and scale-up \citep{glen2018mechanistic,liu2013modelling,lu2015control,nfor2009rational,knappert2020kinetic}.
However, for complex biomanufacturing operations, mechanistic models often oversimplify underlying process mechanisms and make a poor fit to real production data \citep{teixeira2007hybrid}. Thus, 
domain experts are increasingly adopting data-driven methods
for process understanding \citep{mercier2013multivariate,kirdar2007application}, monitoring \citep{teixeira2009situ}, prediction \citep{gunther2009process}, design and control \citep{martagan2017performance}. 

In recent years, various hybrid models were developed to leverage the advantages of mechanistic and data-driven models, such as facilitating bioprocess mechanism learning, and improving prediction, interpretability, and sample efficiency
\citep{lu2015control,solle2017between, vonstoach2014hybrid, von2016hybrid}.
For example, \cite{von2016hybrid} presented a hybrid model that integrates the parametric mechanistic bioreactor model with an artificial neural network. It describes biomass and product formation rates as function of 
fed-batch fermentation conditions.
Hybrid models were further used to support process optimization \citep{zhang2012batch} and scale-up \citep{bollas2003using}. 

\cite{zheng2021policy} proposed a dynamic Bayesian network based hybrid model which can leverage the strengths of  both first-principles and data-driven methods. 
This paper extends such hybrid model 
to account for the key properties of biomanufacturing processes, including nonlinear reactions, nonstationary dynamics, and partially observed state. We
investigate its applicability in cell therapy manufacturing process modeling, prediction, and optimization. 
\end{sloppypar}

\begin{sloppypar}
\textbf{Biomanufacturing Process Optimization:} 
Built on pre-specified process specifications, classic biomanufacturing control strategies tend to 
maintain CPPs within required ranges in order to guarantee product quality \citep{jiang2016integrated} 
through techniques such as feed-forward, feedback control \citep{hong2018challenges,kee2009selective,18cdc_tutorial_control_biology}, and model predictive control \citep{mesbah2017model,Paulson,1383790,lakerveld2013model}. However, 
the existing control strategies derived from PDE/ODE-based mechanistic models often ignore 
process inherent stochasticity and model uncertainty. As most biopharmaceutical manufacturing processes involve significant uncertainties and disturbances, the optimal solution, derived from deterministic mechanistic models, can be suboptimal and result in constraint violations \citep{yoo2021reinforcement}. 


In recent years, RL attracts a lot of attention.
\cite{yoo2021reinforcement} provides a nice review on RL for batch process control. 
To improve sample efficiency, \cite{machalek2021novel} proposed a hybrid machine learning physics models based RL approach.
\cite{zheng2021policy} developed a hybrid model based RL framework that builds the process mechanisms into a Gaussian probabilistic KG, which has been shown to achieve human-level control in low-data environments. However, this method has a strong linearity assumption in the state transition 
model
and fails to take batch-to-batch variation into consideration. To improve the robustness and guarantee the product quality, \cite{pan2021constrained} proposed a constrained Q-learning algorithm 
to meet the specification requirements while \cite{yoo2021reinforcement2} extended the deep deterministic policy gradient algorithm with Monte-Carlo (MC) learning to ensure stable learning behavior.


For a highly variable and complex system with limited data, the combination of Bayesian model inference and RL has attracted increasing attention due to its advantage in handling model uncertainty and inherent stochasticity
\citep{ghavamzadeh2016bayesian}. 
\cite{mowbray2021safe} 
proposed Gaussian processes for data-driven dynamic modelling, which naturally accounts for process stochasticity, and then used it to solve RL-based policy optimization for high probability constraint satisfaction.
\cite{byun2020robust} constructed a dual controller that balances learning and control, i.e., exploration and exploitation.
To provide robustness against uncertainties, the authors formulated a hyper-state that includes a Gaussian belief about uncertain parameters along with the physical states of system as an input state to the control agent. They also designed the reward function to satisfy operational constraints as well as to maximize the final production. The case study of a fed-batch bioehthanol fermentation was presented to show that the proposed controller supports an active learning about uncertain parameters and it can provide better overall performance than conventional certainty equivalence control approaches.
\end{sloppypar}


\section{Problem Statement and Bayesian RL}
\label{sec: problem description}

In this paper, we model the cell therapy manufacturing process as a finite-horizon Markov decision process (MDP) specified by $(\mathcal{S}, \mathcal{A}, H, r, p)$, where $\mathcal{S}$, $\mathcal{A}$, $H$, $r$ and $p$ represent state space, action space, planning horizon, reward function, and state transition probability model. The process state transition probability, accounting for inherent stochasticity and nonstationary dynamics, is modeled as, 
\vspace{-0.in}
\begin{equation*}\label{eq: state transition}
    \pmb{s}_{t+1} \sim p(\pmb{s}_{t+1}|\pmb{s}_t,\pmb{a}_t;\pmb\theta_t)
\end{equation*}
where $\pmb{s}_t\in \mathcal{S}\subset \mathbb{R}^d$ denotes the biomanufacturing process state (e.g., glucose and lactate concentrations, cell density, and CQAs), 
$\pmb{a}_t \in \mathcal{A}$ is the action (also known as control inputs) at time step $t$, $\mathcal{A}$ is a finite set of actions with cardinality $|\mathcal{A}|$, and $t\in\mathcal{H}\equiv\{1,2,\ldots,H\}$ denotes the discrete time index (a.k.a. decision epochs).
The process dynamics and variations are specified by the model parameters, denoted by
$\pmb\theta=(\pmb\theta_1,\ldots,\pmb\theta_H)^\top\in \mathbb{R}^{d_\theta}$. 

\begin{sloppypar}
At any time step $t$,
we take an action by following a policy function mapping the state vector $\pmb{s}_t$ to an action, 
i.e., $\pmb{a}_t = \pi_t\left(\pmb{s}_t\right)$, and collect the reward $r_t(\pmb{s}_t,\pmb{a}_t)$.
Therefore, the probabilistic model of biomanufacturing stochastic decision process (SDP) trajectory $\pmb\tau=(\pmb{s}_1,\pmb{a}_1,\ldots,\pmb{s}_H,\pmb{a}_H,\pmb{s}_{H+1})$, i.e.,
\begin{equation*}
        p(\pmb\tau|{\pmb\theta}) = p(\pmb{s}_1)\sum^{H}_{t=1}p(\pmb{s}_{t+1}|\pmb{s}_t, \pmb{a}_t
        ;{\pmb\theta}_t),
    \label{eq: decision process daynamics}
\end{equation*}
depends on the selection of process model parameters $\pmb{\theta}$ and decision policy $\pi\equiv (\pi_1,\ldots,\pi_H)$.
Given 
any feasible parameters $\pmb{\theta}$, the performance of the policy $\pi$ is evaluated via the expected accumulated reward, 
\begin{equation}
    J\left(\pi;{\pmb \theta}\right) \equiv \E_{\pmb\tau}\left[\left.\sum_{t=1}^{H+1} r_t(\pmb{s}_t,\pmb{a}_t)
\right|\pmb{\pi},
{\pmb{{\theta}}}\right],
\label{eq: objective-simple}
\end{equation}
where $\pmb{s}_{H+1}$ is the terminal state with the reward $r_{H+1}(\pmb{s}_{H+1}, \pmb{a}_{H+1})=
r_{H+1}(\pmb{s}_{H+1})$; for example $\pmb{a}_{H+1}$ is the harvest decision in the cell culture process and the reward only depends on the state $\pmb{s}_{H+1}$.
\end{sloppypar}

\begin{sloppypar}
In the cell therapy manufacturing, there often exist unobservable state variables (such as cell growth inhibitor, metabolic state), which have substantial impact on the cell production and quality. Let $\pmb{z}_t$ denote the latent state variable(s). Thus, at any time step $t$, the process state $\pmb{s}_t$ includes observable and unobservable (latent) state variables, i.e., $\pmb{s}_t=(\pmb{x}_t,\pmb{z}_t)$ with  $\pmb{x}_t\in \mathcal{S}_x$ and latent state variables $\pmb{z}_t\in \mathcal{S}_z$, 
where $\mathcal{S}_x\subset \mathbb{R}^{d_x}$ and $\mathcal{S}_z\subset \mathbb{R}^{d_z}$ with $\mathcal{S}=\mathcal{S}_x\times \mathcal{S}_z$ and $d=d_x+d_z$. By integrating out latent states $(\pmb{z}_1,\ldots,\pmb{z}_{H+1})$, 
we have the likelihood of the partially observed trajectory $\pmb\tau_x\equiv (\pmb{x}_1,\pmb{a}_1,\ldots,\pmb{x}_H,\pmb{a}_H,\pmb{x}_{H+1})$, i.e.,
\begin{equation}
    p(\pmb{\tau}_x| \pmb\theta)= \int p(\pmb\tau|\pmb\theta) \d \pmb{z}_1 \cdots \d \pmb{z}_{H+1}.
    \label{eq.likelihood}
\end{equation}
\end{sloppypar}


\begin{sloppypar}
The unknown KG hybrid model parameters can be estimated by using real-world data, say the process observations with size $m$, denoted by $\mathcal{D}=\{\pmb\tau_x^{(i)};i=1,\ldots,m \}$. In this paper, the \textit{model uncertainty} is quantified by a posterior distribution obtained by applying Bayesian rule,
\begin{equation}
p(\pmb\theta|\mathcal{D}) 
\propto p(\pmb{\theta}) P(\mathcal{D}|\pmb{\theta})
=p(\pmb\theta) \prod_{i=1}^m p\left(\pmb{\tau}_x^{(i)}| \pmb\theta \right)
\label{eq.posterior}
\end{equation}
where the prior $p(\pmb\theta)$ can be used to incorporate the existing knowledge on the model parameters.


The \textit{Bayesian KG}, defined as the integration of the bioprocess probabilistic KG hybrid model and the posterior distribution based model uncertainty quantification, is used to support long-term and reliable predictions. Specifically, the posterior predictive distribution is used to provide any $h$ time periods look-ahead prediction,
\begin{equation}
p(\pmb{x}_{t+h} | \pmb{x}_t, \mathcal{D}) = 
\int p(\pmb{x}_{t+h} | \pmb{x}_t, \pmb{\theta}) 
p(\pmb\theta|\mathcal{D}) d \pmb{\theta},
\label{eq.prediction}
\end{equation}
with $(t+h) \leq H+1$. The prediction risk accounts for both inherent stochasticity through $p(\pmb{x}_{t+h} | \pmb{x}_t, \pmb{\theta})$ and model uncertainty through $p(\pmb\theta|\mathcal{D})$.
Given any $\pmb{\theta}$, the conditional distribution $p(\pmb{x}_{t+h} | \pmb{x}_t, \pmb{\theta})$ is derived based on the probabilistic KG hybrid model characterizing the spatial-temporal causal interdependencies of CPPs/CQAs.

\textit{Therefore, in this paper, we propose a KG hybrid model-based Bayesian RL to find the optimal, robust, and interpretable policy to simultaneously hedge against both bioprocess inherent stochasticity and model uncertainty,} i.e., 
\begin{equation}\label{eq:optimalpolicy}
\pi^\star = \arg\max_{\pi\in\mathcal{P}} \mathcal{J}\left(\pi\right)
\end{equation}
with $\mathcal{P}$ representing the feasible set of decision policy. 
Based on  
the expected accumulated reward $J\left(\pi;{\pmb \theta}\right)$ for any given model parameters $\pmb{\theta}$ presented
in
(\ref{eq: objective-simple}), 
we define the optimization objective for Bayesian RL as
\begin{equation*}\label{eq:scriptJ}
\mathcal{J}\left(\pi\right) \equiv \E_{\pmb{\theta}\sim p\left(\pmb \theta|\mathcal{D}\right)} \left[ J\left(\pi;\pmb \theta\right) \right]
\end{equation*}
with (1) inner expectation in $J(\pi;{\pmb \theta}) = \E_{\pmb\tau}[\sum_{t=1}^{H+1} r_t(\pmb{s}_t,\pmb{a}_t)
|\pmb{\pi},
{\pmb{{\theta}}}]$ accounting for inherent stochasticity; and (2) outer expectation accounting for model uncertainty. 
Since there are often very limited process observations in cell therapy manufacturing especially for personalized bio-drugs,  model uncertainty can be large. Ignoring any source of uncertainty can lead to unreliable and sub-optimal decision making, as well as impacting on the bioprocess mechanism learning through Bayesian inference on the probabilistic KG hybrid model.

\end{sloppypar}


\section{Bioprocess Hybrid Modeling and Inference}
\label{sec:hybridModeling}

\begin{sloppypar}
In this section, we develop a dynamic Bayesian network based hybrid model characterizing the spatial-temporal causal interdependencies of bioprocessing and CPPs/CQAs, such as how decision strategies at different times interact with each other and influence on the production process trajectory dynamics and variation. It can leverage the information from existing mechanistic models from different phases and operation units, and further facilitate learning from process data. 
The PDE/ODE-based mechanistic models typically ignore inherent stochasticity, including batch-to-batch variation, raw material uncertainty, 
and bioprocess noise, which are often dominant sources of biomanufacturing process variation \citep{mockus2015batch}. 
Thus, we will consider bioprocess noise and raw material uncertainty in Section~\ref{subsec:generalHybridModeling}, and further extend the hybrid model to incorporate batch-to-batch kinetics variation in Section~\ref{subsec: b2b variation}.
\textit{The proposed KG hybrid model can provide the risk- and science-based understanding of integrated biomanufacturing mechanisms.} 
It can improve prediction accuracy and reliability.

Since the state transition probabilistic model for $p(\pmb{s}_{t+1}|\pmb{s}_t,\pmb{a}_t;\pmb\theta_t)$ involves latent state variables, nonlinear reactions, and batch-to-batch variation in kinetic coefficients, 
it is challenging to calculate the likelihood of partially observed trajectory in (\ref{eq.likelihood}) and derive the closed form posterior distribution $p(\pmb{\theta}|\mathcal{D})$ by applying Bayesian rule in (\ref{eq.posterior}). 
Thus, in Section~\ref{subsec:BayesianInference}, 
we provide a computational Bayesian inference approach to approximate $p(\pmb{\theta}|\mathcal{D})$ and generate $B$ posterior samples quantifying model uncertainty.

\end{sloppypar}

\subsection{Bioprocess KG Hybrid Model Development}
\label{subsec:generalHybridModeling}

\begin{sloppy}

\end{sloppy}

\begin{sloppy}
Given the existing nonlinear ODE-based mechanistic model, represented by  
${\mbox{d}\pmb{s}}/{\mbox{d}t} = \pmb{f}\left(\pmb{s},\pmb{a}; \pmb\beta\right),$
by using the finite difference approximations for derivatives, i.e., $\mbox{d} \pmb{s}\approx \Delta \pmb{s}_t=\pmb{s}_{t+1}-\pmb{s}_t$, 
and $\mbox{d}t\approx \Delta t$, we construct the hybrid model for state transition, 
\begin{align}
    \pmb{x}_{t+1} &= \pmb{x}_t + \Delta t \cdot \pmb{f}_x(\pmb{x}_t,\pmb{z}_t,\pmb{a}_t; \pmb{\beta}_t) + \pmb{e}^{x}_{t+1}, \label{equ:hybridobs} \\
    \pmb{z}_{t+1} &= \pmb{z}_t + \Delta t \cdot \pmb{f}_z(\pmb{x}_t,\pmb{z}_t,\pmb{a}_t; \pmb{\beta}_t) + \pmb{e}^{z}_{t+1}
    \label{equ:hybridlatent},
\end{align}
with unknown $d_\beta$-dimensional kinetic coefficients $\pmb{\beta}_t\in \mathbb{R}^{d_\beta}$ (e.g., cell growth and inhibition rates). The function structures of $\pmb{f}_x(\cdot)$ and $\pmb{f}_z(\cdot)$ are derived from $\pmb{f}(\cdot)$ in the mechanistic models.
The residual terms are modeled by  multivariate Gaussian distributions $\pmb{e}_{t+1}^{x} \sim \mathcal{N}(0,V^{x}_{t+1})$ and $\pmb{e}_{t+1}^{z} \sim \mathcal{N}(0,V^{z}_{t+1})$ with zero means and covariance matrices $V^{x}_{t+1}$ and $V^{z}_{t+1}$. 
The kinetic coefficients $\pmb\beta_t$ can change over different phases of bioprocess (such as growth and stationary phases in cell culture process) to represent the fact that the bioprocess dynamics can be time-varying. This facilitates a modular and flexible
design of integrated biomanufacturing processes.
For notation simplification, we consider fixed time step $\Delta t$ even through the proposed methodology can be applicable to general situations with heterogenous process data integrating both online and offline measuremements.

{The statistical residual terms $\pmb{e}_t=(\pmb{e}_t^x,\pmb{e}_t^z)$ allow us to account for the impact from bioprocess noise, raw material uncertainty, 
ignored critical process parameters (CPPs), and other uncontrollable factors (e.g., contamination) occurring at any time step $t$.} Thus, this KG hybrid model, integrating with Bayesian model inference, can facilitate mechanism learning, guide the production process variation reduction, and support Quality-by-Design (QbD) and robust control.

Let $\pmb{g}_x(\pmb{x}_t, \pmb{z}_t, \pmb{a}_t; \pmb{\beta}_t)\equiv\pmb{x}_t + \Delta t \cdot \pmb{f}_x(\pmb{x}_t,\pmb{z}_t,\pmb{a}_t; \pmb{\beta}_t)$ and $\pmb{g}_z(\pmb{x}_t, \pmb{z}_t, \pmb{a}_t; \pmb{\beta}_t)\equiv\pmb{z}_t + \Delta t \cdot \pmb{f}_z(\pmb{x}_t,\pmb{z}_t,\pmb{a}_t; \pmb{\beta}_t)$.
Equivalently, we can rewrite the state transition model \eqref{equ:hybridobs} and \eqref{equ:hybridlatent} as,
\begin{align*}
    & \pmb{x}_{t+1}|\pmb{x}_t,\pmb{z}_t,\pmb{a}_t 
    \sim \mathcal{N}\Big(\pmb{g}_x(\pmb{x}_t, \pmb{z}_t, \pmb{a}_t; \pmb{\beta}_t),  V_{t+1}^{\pmb{x}} \Big)
    \\
    & \pmb{z}_{t+1}|\pmb{x}_t,\pmb{z}_t,\pmb{a}_t 
    \sim  \mathcal{N}\Big(\pmb{g}_z(\pmb{x}_t, \pmb{z}_t, \pmb{a}_t; \pmb{\beta}_t),  V_{t+1}^{\pmb{z}} \Big). 
\end{align*}
Let the state transition model parameters $\pmb\theta_t$ representing the composite parameters including kinetic coefficients and the variances of process residuals, i.e.,
$\pmb{\theta}_t=(\pmb\beta_t,\mbox{vec}(V^x_{t+1}),\mbox{vec}(V^z_{t+1}))^\top$. We use $\text{vec}\left(\cdot\right)$ to denote a linear transformation converting an $n\times m$ matrix into a $nm\times 1$ column vector.
Let $g(\pmb{s}_t,\pmb{a}_t;\pmb{\beta}_t) \equiv \pmb{s}_t + \Delta t \cdot \pmb{f}(\pmb{s}_t,\pmb{a}_t; \pmb{\beta}_t)$. Then, at any time step $t\in\mathcal{H}$, we have the full state transition
\begin{align}
   \pmb{s}_{t+1}|\pmb{s}_{t},\pmb{a}_t
   &= g(\pmb{s}_t,\pmb{a}_t;\pmb{\beta}_t) + \pmb{e}_{t+1} \nonumber\\ &\sim \mathcal{N}\Big(g(\pmb{s}_t,\pmb{a}_t;\pmb{\beta}_t),  V_{t+1} \Big)\label{eq: composite state transition}
\end{align}
where $\pmb{s}_t=(\pmb{x}_t,\pmb{z}_t)$, $\pmb{e}_{t+1}=(\pmb{e}^x_{t+1},\pmb{e}^z_{t+1})$, and $V_{t+1}$ is diagonal covariance matrix with diagonal entries from $V^x_{t+1}$ and $V^z_{t+1}$.
\end{sloppy}

\begin{figure}[ht]
	\centering
 	\includegraphics[width=0.5\textwidth]{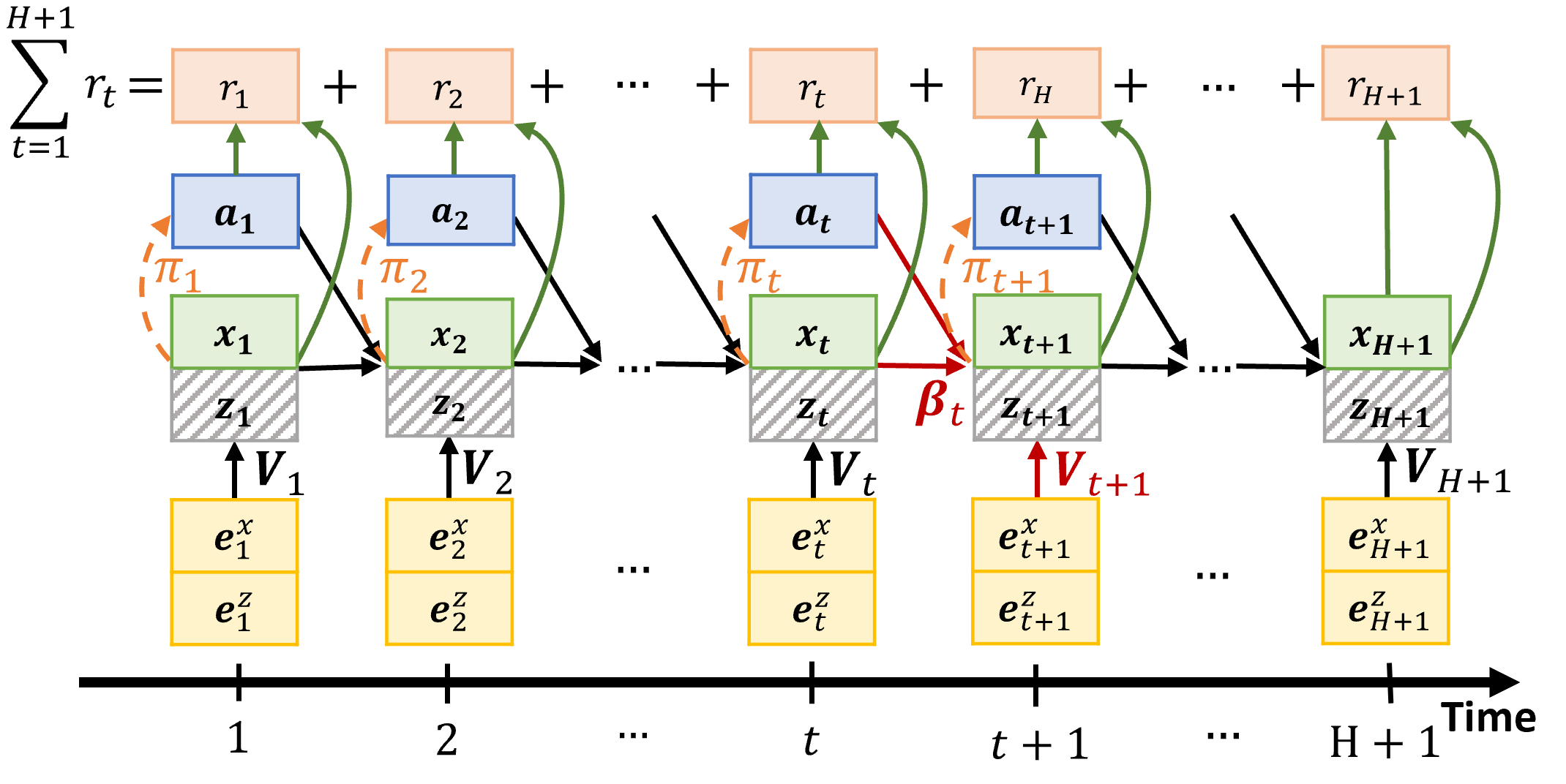}
 	\caption{An illustration of policy-augmented knowledge graph (KG) network for the stochastic decision process hybrid model.} \label{fig:hybrid_example}
\end{figure}

\begin{sloppypar}
The probabilistic KG
of integrated biomanufacturing process can be visualized by a directed network as shown in Figure~\ref{fig:hybrid_example}.
The observed state $\pmb{x}_t$ and latent state $\pmb{z}_t$ are represented by solid and shaded nodes respectively. The directed edges represent causal interactions.
At any time period $t+1$, the process state output node
$\pmb{s}_{t+1}=(\pmb{x}_{t+1},\pmb{z}_{t+1})$ depends on its parent nodes: $\pmb{s}_{t+1}=\pmb{f}(Pa(\pmb{s}_{t+1});{\pmb\theta}_t)$
with $Pa(\pmb{s}_{t+1})=(\pmb{s}_t,\pmb{a}_t,\pmb{e}_{t+1})$ and $\pmb{\theta}_t=(\pmb{\beta}_t,{V}_{t+1})$.
We create a \textit{policy augmented KG network} by including additional edges: 1) connecting state $\pmb{s}_t$ to action $\pmb{a}_t$ representing the causal effect of the policy, $\pmb{a}_t=\pi_t (\pmb{s}_t)$; and 2) connecting actions and states to the immediate reward $r_t(\pmb{s}_t,\pmb{a}_t)$  (e.g., cost, cell product productivity).
This network models how the effect of current state and action, $\{\pmb{s}_t,\pmb{a}_t\}$, propagates through mechanism pathways impacting on the output trajectory and the accumulated reward. 
\textit{Thus, the proposed Bayesian KG built in conjunction with RL can support long-term prediction and guide interpretable, reliable, and optimal decision making.}
\end{sloppypar}

In this paper, we consider case studies with some prior knowledge on the underlying bioprocess mechanisms --- the form of function $\pmb{f}(\cdot)$ is known but the kinetics coefficients $\pmb\beta$ are unknown. In the situations 
without any strong prior knowledge
on the underlying bioprocess mechanisms (i.e., $\pmb{f}(\cdot)$ is unknown) or for the cases with intensive online monitoring 
(i.e., $\Delta t$ is small), for example in-situ Raman spectroscopy probe is used to real-time measure cell density and the concentrations of nutrients and metabolites in the bioreactors \citep{mehdizadeh2015generic,craven2014glucose}, we can use a linear statistical function modeling the state transition to consider the main effects. Specifically, at any time step $t\in\mathcal{H}$, we model the state transition as $\pmb{s}_{t+1}= \pmb{\beta}^{\pmb{s}}_t  \pmb{s}_t + \pmb{\beta}^{\pmb{a}}_t  \pmb{a}_t + \pmb{e}_{t+1}$, where the coefficients $\pmb{\beta}^{\pmb{s}}_t$ and $\pmb{\beta}^{\pmb{a}}_t$ measure the main effects of current state $\pmb{s}_t$ and action $\pmb{a}_t$ on next state $\pmb{s}_{t+1}$, and $\pmb{e}_{t+1}$ represents the residual. 
We refer to the studies \citep{zheng2021policy,xie2020interpretable} for more details.


\subsection{Batch-to-Batch Variation Modeling}
\label{subsec: b2b variation}


\begin{sloppypar}

To further account for batch-to-batch variation, 
the kinetic coefficients are modeled by random variables,
\begin{equation*}\label{eq: batch-to-batch var model}
    \pmb\beta_t=\pmb\mu_t^\beta+\pmb\epsilon_t^\beta
\end{equation*}
where $\pmb\epsilon^\beta_t$ represents the random batch effect following a multivariate normal distribution with mean zero
and covariance matrix $\Sigma^\beta$. The vector $\pmb\mu_t^\beta$ represents the mean of kinetic coefficients.
Then, the state transition in \eqref{equ:hybridobs} and \eqref{equ:hybridlatent} becomes
\begin{align}
    \pmb{x}_{t+1} &=\pmb{g}_x(\pmb{x}_t, \pmb{z}_t, \pmb{a}_t, \pmb{\beta}_t; \pmb{\theta}_t) + \pmb{e}^{x}_{t+1}, \label{equ:hybridobs-b2b} \\
    \pmb{z}_{t+1} &= \pmb{g}_z(\pmb{x}_t, \pmb{z}_t, \pmb{a}_t, \pmb{\beta}_t; \pmb{\theta}_t) + \pmb{e}^{z}_{t+1}
    \label{equ:hybridlatent-b2b},
\end{align}
where $\pmb\beta_t\sim \mathcal{N}(\pmb\mu_t^\beta,\Sigma^\beta_t)$, $\pmb{e}^{x}_{t+1}\sim \mathcal{N}(0,V^x_{t+1})$, and $\pmb{e}^{z}_{t+1}\sim \mathcal{N}(0,V^z_{t+1})$ represent inherent stochasticity.
Therefore, the biomanufacturing KG hybrid model is specified by the parameters,
$\pmb{\theta}_t=\big(\pmb\mu_t^\beta, \mbox{vec}(\Sigma_t^\beta),\mbox{vec}(V^x_{t+1}),\mbox{vec}(V^z_{t+1})\big)^\top.$ 
\end{sloppypar}

\subsection{KG Hybrid Model Uncertainty Quantification}
\label{subsec:BayesianInference}

\begin{sloppy}

\end{sloppy}

\begin{sloppypar}
Given process observations $\mathcal{D}=\{\pmb\tau_x^{(i)};i=1,\ldots,m\}$, the approximate Bayesian computation (ABC) approach \citep{minter2019approximate,toni2009approximate,mckinley2018approximate,beaumont2009adaptive} is used to generate posterior samples  
quantifying the KG hybrid model parameter estimation uncertainty. 
For each $i$-th process trajectory observation $\pmb\tau_x^{(i)}$, the algorithm jointly simulates model parameters $\pmb{\theta} \sim p(\pmb{\theta})$ and predicted trajectories $\pmb\tau_x^{\star(ij)} \sim p(\pmb\tau_x|\pmb{\theta})$ for $j=1,2\ldots, L$. We accept the samples of parameters (called particles) when the averaged distance between the real process observations and simulated data is under a certain tolerance level denoted by $\delta>0$. We define the distance as,
\begin{equation*}
    q\equiv\frac{1}{mL}\sum^m_{i=1}\sum^L_{j=1}d(\pmb\tau_x^{(i)},\pmb\tau_x^{\star(ij)}), 
\end{equation*} where $d(\cdot,\cdot)$ is the distance function and  $\pmb\tau^{\star(i\cdot)}_x$ is the predicted partially observed process trajectory. 
Here we use the square root of the sum of squared differences between trajectories as the distance measure, i.e., $d\left(\pmb{x},\pmb{y}\right) \equiv \left\Vert\pmb{x} - \pmb{y}\right\Vert$ for any $\pmb{x}$ and $\pmb{y}$. For any $d$-dimensional vector $\pmb{x}\in \mathbb{R}^d$, we consider L2 norm, i.e., $\Vert\pmb{x}\Vert 
=\sqrt{\sum_{i=1}^d x_i^2}$.
Through gradually reducing the tolerance level $\delta$, we can better approximate the posterior distribution $p(\pmb{\theta}|\mathcal{D})$
with $p(\pmb{\theta}|q \leq \delta).$

\end{sloppypar}

\begin{sloppypar}
We use the sequential Monte Carlo (SMC) to efficiently explore the parameter space and generate posterior samples. 
The procedure of the ABC-SMC sampling approach is summarized at {Algorithm~\ref{Algr:SMC-ABC}}.
The initial set of parameter samples $\{\pmb{{\theta}}_{n}^{(0)}\}_{n=1}^N$ is generated from the prior distribution $p({\pmb{\theta}})$ in {Step~1}. 
The associated weights $\{w_{n}^{(0)}\}_{n=1}^N$ and distances $\{q_{n}^{(0)}\}_{n=1}^N$ are calculated in {Steps~2-3}. 
For each $i$-th observation, we generate $L$ predicted trajectories and calculate the distance $q_n^{(0)}$. 
The tolerance level $\delta_g$ in the first iteration $g=1$ is determined online as the $\alpha$-quantile of the $\{q_n^{(0)}\}_{n = 1}^{N}$ in {Step~4}. Particles $\{\pmb{\theta}_n\}_{n=1}^{N_\alpha}$ satisfying this tolerance constitute the weighted empirical distribution to approximate the posterior distribution in {Step~5}, where $N_\alpha = \lfloor \alpha N \rfloor$. 
The approximation accuracy is measured by the corresponding distances $\{q_{n}\}_{n=1}^{N_\alpha}$. 
Then, $N-N_\alpha$ new particles are drawn from the sample set from the previous $(g-1)$-th iteration, i.e., $\{\pmb{{\theta}}_k^{g-1}\}_{k=1}^{N_\alpha}$, with corresponding weights and perturbed by the kernel function $K$ in {Steps~6-7}. Similar to {Steps~1-5}, the associated weights and distances are calculated in {Steps~8-10}, the tolerance level $\delta_g$ and the posterior distribution approximate are updated in {Steps 12-13}. We repeat Steps~6-13 until the proportion 
of particles satisfying the tolerance level $\delta_{g-1}$ among the $N-N_\alpha$ new particles is below the pre-specified threshold $p_{acc_{min}}$. Finally, the algorithm returns the weighted empirical distribution as posterior distribution approximate, 
\begin{equation*}
    \widehat{p}(\pmb{\theta}|\mathcal{D}) = \frac{1}{\sum_{n'=1}^{N_\alpha} w_{n'}^{(g-1)}} \sum_{n=1}^{N_\alpha} w_n^{(g-1)} \mathbb{1}\left(\pmb{\theta} = \pmb{\theta}_n^{(g-1)}\right).
\end{equation*} 

\begin{algorithm}[ht] 
\DontPrintSemicolon
\KwIn{
the prior distribution $p(\pmb{\theta})$; the number of particles $N$; 
process observations $\{\pmb\tau_x^{(i)}\}_{i=1}^m$;  the perturbation kernel function $K(\cdot)$; the number of particles to keep at each iteration 
$N_\alpha = \lfloor \alpha N \rfloor$ with $\alpha \in [0,1]$; and the minimal acceptance rate $p_{acc_{min}}$.
}
\KwOut{posterior distribution approximate $\widehat{p}(\pmb{\theta}|\mathcal{D})$.
 } 
{   
    \For{$n = 1,\ldots, N$}{
    \textbf{1.} Sample $\pmb{\theta}^{(0)}_n\sim p(\pmb{\theta})$; \\
    \textbf{2.} 
    Generate $L$ predicted trajectories $\{\pmb\tau_{x}^{\star(ij)}\}_{j=1}^L$ using $\pmb{\theta}^{(0)}_n$ with $i=1,\ldots,m$;\\ 
    \textbf{3.} Set $q_n^{(0)} = \frac{1}{mL}\sum_{i=1}^{m}\sum_{j=1}^{L}  d(\pmb\tau_{x}^{(i)}, \pmb\tau_{x}^{\star(ij)})$ and $w_n^{(0)} = 1$;
    }
    \textbf{4.} Let $\delta_1$ 
    be the first $\alpha$-quantile of $q^{(0)} = \{q_n^{(0)}\}_{n = 1}^{N}$;\\
    \textbf{5.} Let $\{(\pmb{\theta}_n^{(1)},w_n^{(1)},q_n^{(1)})\} = \{(\pmb{\theta}_n^{(0)},w_n^{(0)},q_n^{(0)})|q_n^{(0)} \leq \delta_1, 1\leq n \leq N\}$, $p_{acc} = 1$ and $g = 2$;\\
    \While{$p_{acc} > p_{acc_{min}}$}{
        \For{$n = N_\alpha + 1,\ldots,N$}{
            \textbf{6.} Sample $\pmb{\theta}^\star_n$ from $\pmb\theta^{(g-1)}_k$ with probability $\frac{w^{(g-1)}_k}{\sum_{j=1}^{N_\alpha}w_j^{(g-1)}}$, $1 \leq k \leq N_\alpha$;\\
            \textbf{7.} Perturb the particle to obtain $\pmb{\theta}^{(g-1)}_n \sim K(\pmb{\theta}|\pmb{\theta}^{\star}_n) = \mathcal{N}(\pmb{\theta}^{\star}_n,\sum)$; \\
            \textbf{8.} 
            Generate $L$ predicted trajectories $\{\pmb\tau_{x}^{\star(ij)}\}_{j=1}^L$ using $\pmb{\theta}^{(g-1)}_n$ with $i=1,\ldots,m$; \\
            \textbf{9.} Set $q_n^{(g-1)} = \frac{1}{mL}\sum_{i=1}^{m}\sum_{j=1}^{L}  d(\pmb\tau_{x}^{(i)}, \pmb\tau_{x}^{\star(ij)})$;\\
            \textbf{10.} Set ${w}_n^{(g-1)} =   \frac{p(\pmb{\theta}^{(g-1)}_n) \sum_{i=1}^{m}\sum_{j=1}^{L}  \mathbb{1}(d(\pmb\tau_{x}^{(i)}, \pmb\tau_{x}^{\star(ij)} \leq \delta_{g-1})}{\sum_{j=1}^{N_\alpha} \frac{w_j^{(g-1)}}{ \sum_{k=1}^{N_\alpha} w_k^{(g-1)}} K(\pmb{\theta}_n^{(g-1)}|\pmb{\theta}^{(g-1)}_{j})}$;
        }
    \textbf{11.} Set $p_{acc} = \frac{1}{N-N_\alpha}\sum_{k=N_\alpha+1}^N \mathbb{1}(q_k^{(g-1)} \leq \delta_{g-1})$; \\
    \textbf{12.} Let $\delta_g$ 
    be the first $\alpha$-quantile of 
    $q^{(g-1)} = \{q_n^{(g-1)}\}_{n = 1}^{N}$;\\
    \textbf{13.} Let $\{(\pmb{\theta}_n^{(g)},w_n^{(g)},q_n^{(g)})\} = \{(\pmb{\theta}_n^{(g-1)},w_n^{(g-1)},q_n^{(g-1)})|q_n^{(g-1)} \leq \delta_{g}, 1\leq n \leq N\}$ and $g = g + 1$;\\
    }
    \textbf{14. Return} the approximated posterior distribution, $\widehat{p}(\pmb{\theta}|\mathcal{D}) = \frac{1}{\sum_{n'=1}^{N_\alpha} w^{(g-1)}_{n'}} \sum_{n=1}^{N_\alpha} w^{(g-1)}_n \mathbb{1}\left(\pmb{\theta} = \pmb{\theta}^{(g-1)}_n  \right)$.
}
\caption{ABC-SMC sampling procedure for generating posterior samples from $p(\pmb{\theta}|\mathcal{D})$ (derived from \cite{toni2009approximate,lenormand2013adaptive,del2006sequential})}
\label{Algr:SMC-ABC}
\end{algorithm}

\end{sloppypar}
\section{KG Hybrid Model-based Bayesian RL}
\label{sec:RL}




\begin{sloppypar}
We formulate the biomanufacturing process optimization and control as a finite-horizon MDP.
The process starts with an initial state $\pmb{s}_1$. 
At any time step $t$, the agent observes the state $\pmb{s}_t$, chooses an action $\pmb{a}_t$, 
and receives a reward $r_t = r_t(\pmb{s}_t, \pmb{a}_t)$. 
Then the process transits to a new state $\pmb{s}_{t+1}$ at next time step $t+1$ by following the state transition distribution $p(\pmb{s}_{t+1}|\pmb{s}_t,\pmb{a}_t;\pmb{\theta}_t)$. 
The episode terminates when $\pmb{s}_{H+1}$ is reached, such as 
the agent takes the ``harvest'' action in the cell culture process, and collect the reward $r_{H+1}(\pmb{s}_{H+1})$.
Since there are often very limited process observations in cell therapy manufacturing, the goal is to find a \textit{sample efficiently} optimization approach to learn the optimal policy in \eqref{eq:optimalpolicy} 
accounting for both inherent stochasticity and model uncertainty. 
\end{sloppypar}

\begin{sloppypar}
For each $t\in \mathcal{H}$, we define the value function $V_t^{\pi}(\pmb{s}): \mathcal{S}\rightarrow \mathbb{R}$ as the expected value of cumulative rewards received under policy $\pi$ when starting from state $\pmb{s}$ 
at the $t$-th step, i.e.,
\begin{equation*}
    V_t^\pi(\pmb{s}) = \E_{ p(\pmb\theta|\mathcal{D})}\E_{p(\pmb{s}_{t+1}|\pmb{s}_t,\pi_t(\pmb{s}_t);\pmb{\theta}_t)}\left[\left.\sum^H_{\ell=t} r_\ell(\pmb{s}_\ell,\pi_\ell(\pmb{s}_\ell)) \right| \pmb{s}_t=\pmb{s}\right]
\end{equation*}
where the outer expectation $\E_{ p(\pmb\theta|\mathcal{D})}[\cdot]$ accounts for model uncertainty. Given any posterior sample of model parameters ${\pmb{\theta}}\sim p(\pmb{\theta}|\mathcal{D})$, the inner expectation $\E_{p(\pmb{s}_{t+1}|\pmb{s}_t,\pi(\pmb{s}_t);\pmb{\theta}_t)}[\cdot]$ is taken over the state transition model \eqref{equ:hybridobs-b2b} and \eqref{equ:hybridlatent-b2b} under the policy function $\pi$.
 Accordingly, we also define the action-value Q-function, denoted by $Q^\pi_t :\mathcal{S}\times \mathcal{A} \rightarrow \mathbb{R}$. It gives the expected value
of cumulative rewards when the agent starts from any feasible state-action pair, say $(\pmb{s},\pmb{a})\in\mathcal{S}\times \mathcal{A}$, at the $t$-th step and follows
policy $\pi$ afterwards, i.e., 
\begin{eqnarray}
\lefteqn{Q_t^\pi(\pmb{s}, \pmb{a}) = \E\left[\left.\sum^H_{\ell=t}  r_\ell(\pmb{s}_\ell,\pi_\ell(\pmb{s}_\ell)) \right| \pmb{s}_t =\pmb{s},\pmb{a}_t=\pmb{a}\right]} \nonumber\\
&&= r_t(\pmb s,\pmb a) + \E\left[\left.V^\pi_{t+1}(\pmb{s}_{t+1})\right| \pmb{s}_t=\pmb{s},\pmb{a}_t=\pmb{a}\right] \hspace{0.13in}
\label{eq: Q value definition}
\end{eqnarray}
where the expectation is taken over both posterior distribution $p(\pmb{\theta}|\mathcal{D})$ and 
the transition distribution $p(\pmb{s}_{t+1}| \pmb{s}_t, \pmb{a}_t;\pmb\theta_t)$ for all $t\in T$ by following the policy $\pmb{a}_t =\pi_t(\pmb{s}_t)$. For notation simplification, we use $\E[\cdot]=\E_{ p(\pmb\theta|\mathcal{D})}\E_{p(\pmb{s}_{t+1}|\pmb{s}_t,\pi_t(\pmb{s}_t);\pmb{\theta}_t)}\left[\cdot\right]$ in the following discussion if not mentioned otherwise.

Eq.~\eqref{eq: Q value definition} is known as the Bellman equation which expresses a relationship between the value of a state and the values of its successor states \citep{sutton2018reinforcement}. Considering the optimal value function $V_t^\star(\pmb{s}) = \max_{\pi\in \mathcal{P}}V_t^\pi(\pmb{s})$, we have the Bellman optimality equation \citep[Chapter~3.6]{sutton2018reinforcement}
\begin{align}
V_t^\star(\pmb{s})&=\max_{\pmb{a}\in\mathcal{A}} r_t(\pmb s,\pmb a) + \E\left[\left.V^\star_{t+1}(\pmb{s}_{t+1})\right| \pmb{s}_t=\pmb{s},\pmb{a}_t=\pmb{a}\right]\nonumber\\
&=\max_{\pmb{a}\in\mathcal{A}} Q_t^\star(\pmb{s}, \pmb{a}).
\label{eq: Bellman optimality equation}
\end{align}
We use the finite sampling approximation (SAA) to estimate the expectation in  \eqref{eq: Bellman optimality equation} 
during the optimal Q-function and value-function estimation in Algorithm~\ref{Algr: Bayesian sparse sampling}.
It is known that the optimal policy $\pi^\star$ is the greedy policy with respect to the optimal Q-value function, $Q_t^\star(\pmb{s},\pmb{a}) = \max_{\pi\in \mathcal{P}}Q_t^\pi(\pmb{s},\pmb{a})$ \citep[Chapter~3.8]{sutton2018reinforcement}. Thus, 
we 
obtain the optimal greedy policy \citep{puterman2014markov} with
\begin{equation}\label{eq: greedy policy}
    \pi^\star_t(\pmb{s})\equiv\argmax_{\pmb{a}\in\mathcal{A}}Q^\star_t(\pmb{s},\pmb{a}), \text{ for any $\pmb{s}\in\mathcal{S}$}.
\end{equation}
\end{sloppypar}

\begin{algorithm} 
\DontPrintSemicolon
\KwIn{Current state $\pmb{s}_t$; scenario numbers $B$ and $J$ for the SAA on 
$\E_{ p(\pmb\theta|\mathcal{D})}
\E_{p(\pmb{s}_{t+1}|\pmb{s}_t,\pi(\pmb{s}_t);\pmb{\theta}_t)}[\cdot]$;
posterior $\widehat{p}(\pmb{\theta}|\mathcal{D})$ from Algorithm~\ref{Algr:SMC-ABC}.
}
\KwOut{Estimated optimal Q-function $\widehat{Q}(\pmb{s},\pmb{a})$}
    \SetKwProg{Fn}{Function}{:}{}
    \Fn{\textproc{Qfun}{$(t,\pmb{s}_t,\pmb{a}_t)$}}{
            \For{$b=1,2\ldots,B$}{
                \textbf{(A1)} Generate a posterior sample of model parameters, 
                ${\pmb{\theta}}_{b}\sim \widehat{p}(\pmb\theta_t|\mathcal{D})$.
                \\
                \For{$j = 1,\ldots,J$}{
                \textbf{(A2)} Sample from state transition distribution, $\pmb{s}_{t+1}^{(b,j)}\sim p(\pmb{s}_{t+1}|\pmb{s}_t,\pmb{a}_t;{\pmb{\theta}}_{t,b})$\\
                \textbf{(A3)} $V_{t+1}\left(\pmb{s}_{t+1}^{(b,j)}\right)= \textproc{Vfun}\left(t+1,\pmb{s}_{t+1}^{(b,j)}\right)$
                
                }
            }
            \textbf{(A4)} $\widehat{Q}_t(\pmb{s}_t,\pmb{a}_t)=r_t(\pmb{s}_t,\pmb{a}_t)+\frac{1}{BJ}\sum^B_{b=1}\sum^J_{j=1} V_{t+1}\left(\pmb{s}_{t+1}^{(b,j)}\right). $ \\
        \textbf{return} $ \hat{Q}_t(\pmb{s}_t,\pmb{a}_t). $ 
}
\textbf{End Function}

    \Fn{\textproc{Vfun}{$(t,\pmb{s}_t)$}}{
    \If{$t=H+1$}{
    \textbf{return} $r_{H+1}(\pmb{s}_{H+1})$; 
    }
        \For{$\pmb{a}_t\in \mathcal{A}$}
            {
            \For{$b=1,2\ldots,B$}{
                \textbf{(B1)} Generate a posterior sample of model parameters ${\pmb{\theta}}_{b}\sim \widehat{p}(\pmb\theta_t|\mathcal{D})$\\
                \For{$j = 1,\ldots,J$}{
                \textbf{(B2)} Sample from state transition $\pmb{s}_{t+1}^{(b,j)}\sim p(\pmb{s}_{t+1}|\pmb{s}_t,\pmb{a}_t;{\pmb{\theta}}_{t,b})$\\
                \textbf{(B3)} $V_{t+1}\left(\pmb{s}_{t+1}^{(b,j)}\right)= \textproc{Vfun}\left(t+1,\pmb{s}_{t+1}^{(b,j)}\right)$
                
                }
            }
             \textbf{(B4)} Estimate $\widehat{Q}_{t}(\pmb{s}_t,\pmb{a}_t)= r_t(\pmb{s}_t,\pmb{a}_t)+\frac{1}{BJ}\sum^B_{b=1}\sum^J_{j=1} V_{t+1}\left(\pmb{s}_{t+1}^{(b,j)}\right)$
            }
         \textbf{(B5)} $\widehat{V}_t(\pmb{s}_t)=\max_{\pmb{a}_t\in\mathcal{A}}\widehat{Q}_{t}(\pmb{s}_t,\pmb{a}_t)$ as in \eqref{eq: Bellman optimality equation}\\
}
\textbf{End Function}
\caption{Bayesian Sparse Sampling for Estimating Optimal Q-Function and Value Function
}
\label{Algr: Bayesian sparse sampling}
\end{algorithm}



\begin{sloppypar}
We estimate the optimal Q-function $Q^\star_t(\pmb{s},\pmb{a})$ by simulating the probabilistic KG hybrid model and applying the sparse sampling method \citep{wang2021optimizing,kearns2002sparse,wang2005bayesian}. 
Specifically, we generate future trajectories accounting for both bioprocess inherent stochasticity and model uncertainty.
Then, we search for the optimal action through look-ahead tree. Bayesian sparse sampling grows a look-ahead tree starting with the root at the current state, enumerating actions at decision nodes, and sampling at outcome nodes over both posterior distribution and state transition distribution. The same procedure is repeated for each node
until all the leaf nodes reach the end of the planning horizon.

The procedure of Bayesian sparse sampling for optimal Q-function and value function estimation is summarized in Algorithm \ref{Algr: Bayesian sparse sampling}. The Q-function (and the value function) estimates are computed from leaf nodes rolling-backing up to the root. To compute the value function \textproc{Vfun}$(t,\pmb{s}_t)$, we grow a look-ahead tree starting with the root at the current state. For each action $\pmb{a}\in\mathcal{A}$, we grow the root into $BJ$ child nodes with each representing a sample of the next state, i.e., $\pmb{s}_{t+1}^{(b,j)}$ from state transition probability $p(\pmb{s}_{t+1}|\pmb{s}_t,\pmb{a}_t;{\pmb{\theta}}_{t,b})$, where the posterior samples ${\pmb{\theta}}_{t,b}\sim \widehat{p}(\pmb{\theta}|\mathcal{D})$ with $b=1,\ldots,B$ are generated with ABC-SMC and we create $J$ trajectories at each model posterior sample  (Steps~{B1}-{B3}). The same procedure is repeated for each node
until all the leaf nodes reach the end of the process. 
Then, the expected Q-value in \eqref{eq: Q value definition} is 
estimated by SAA in {Step} {B4}. The optimal value of current state and the optimal action are obtained by maximizing the Q values ({Step}~{B5}). The Q-value function ``\textproc{Qfun}'' is calculated in a similar way as ``\textproc{Vfun}''. The only difference is that the current action $\pmb{a}_t$ is given as the input of \textproc{Qfun} so that we remove the step of enumerating all candidate actions at time step $t$ and then return $\widehat{Q}_t(\pmb{s}_t,\pmb{a}_t)$ calculated in Step~B4.


The sampling procedure of KG hybrid model-based Bayesian RL for process optimization is summarized in Algorithm~\ref{Algr: optimization}. The goal of process optimization is to find the optimal action $\pmb{a}^\star_t$ at each time step which maximizes the expected total rewards. To solve the problem, we propose a solution approach which estimates the optimal Q-value function \eqref{eq: Q value definition} from starting time $t$ to the end of planning horizon by calling Algorithm~\ref{Algr: Bayesian sparse sampling} in {Step 1}, and then selects the optimal action by using the greedy policy \eqref{eq: greedy policy} in {Step 2}. After that, we execute the action and the process evolves to the next state at which the entire procedure is repeated.  

\end{sloppypar}

\begin{algorithm}[h!]
\DontPrintSemicolon
\KwIn{The initial state $\pmb{s}_1$
}
\KwOut{Optimal actions $\pmb{a}_t^\star$ for $t=1,2,\ldots,H$}
{   
    \For{$t=1,2,\ldots,H$}{
    \For{$\pmb{a}_t\in \mathcal{A}$}{
    \textbf{1.} $q_t(\pmb{s}_t,\pmb{a}_t)=\textproc{Qfun}(t,\pmb{s}_{t},\pmb{a}_t)$ by calling Algorithm~\ref{Algr: Bayesian sparse sampling} to estimate the optimal Q-value;\\
    }
        \textbf{2.} Find the greedy optimal action \eqref{eq: greedy policy}: $\pmb{a}_t^\star=\argmax_{\pmb{a}_t\in\mathcal{A}} q_t(\pmb{s}_t, \pmb{a}_t)$; \\
        \textbf{3.} Apply the optimal action and evolve to next state, i.e., 
        $\pmb{s}_{t+1}=\E_{p(\pmb\theta_{t}|\mathcal{D})}\left[\E[\pmb{s}_{t+1}|\pmb{s}_{t},\pmb{a}_{t}^\star;\pmb\theta_{t}]\right]$.\\
    }
}
\caption{Hybrid model-based Bayesian RL Sampling Procedure for Process Optimization.}
\label{Algr: optimization}
\end{algorithm}






\section{Cell Therapy Manufacturing Case Study}
\label{sec:caseStudy}


In this section, we use the erythroblast cell therapy manufacturing example presented in \cite{glen2018mechanistic} to assess the performance of proposed hybrid model-based Bayesian RL. The cell culture of erythroblast exhibits two phases: a relatively uninhibited growth phase followed by an inhibited phase. 
\cite{glen2018mechanistic} identified that this reversible inhibition is caused by a unknown cell-driven factor rather than commonly known mass transfer or metabolic limitations.

\subsection{Cell Culture Process Hybrid Modeling}
\begin{sloppy}
\cite{glen2018mechanistic} developed an ODE-based mechanistic model describing the dynamics of an unidentified autocrine growth inhibitor  accumulation 
and its impact on the erythroblast cell production process.
We extend this model to a two-phase cell culture process, including growth and stationary phases with index $p=1,2$. 
The mechanistic model of cell growth and inhibitor accumulation 
for the $p$-th phase is
\begin{align}
    \frac{\d\rho}{\d t} &= r^g_p \rho \Bigg (1 - \Big(1+e^{(k^s_p(k^c_p-I))} \Big) ^{-1} \Bigg ),
    \label{equ:ODEcell} \\
    \frac{\d I}{\d t} &= \frac{\d\rho}{\d t} - r^d_p I
    \label{equ:ODEinhibition},
\end{align}
where $\rho_t$ and $I_t$ represent the cell density and the inhibitor concentration at time step $t$.
The kinetic coefficients $r^g_p$, $k^s_p$, $k^c_p$ and $r^d_p$ 
denote the cell growth rate, inhibitor sensitivity, inhibitor threshold, and inhibitor decay.
Suppose that the phase transition 
occurs at $T_\star=18$ hour. 
\end{sloppy}

\begin{sloppypar}
Given the mechanistic model in 
(\ref{equ:ODEcell})--(\ref{equ:ODEinhibition}), by following the derivation in Section~\ref{subsec:generalHybridModeling},
we construct the hybrid model,
\begin{align}
    \rho_{t+1} &= \rho_t + \Delta t \cdot r^g_p \rho_t \Bigg (1 - \Big(1+e^{(k^s_p(k^c_p-I_t))} \Big) ^{-1} \Bigg ) + e^{\rho}_{t},\nonumber\\
    I_{t+1} &= I_t + \Delta t \cdot \Bigg (\frac{\rho_{t+1}-\rho_t}{\Delta t} - r^d_p I_t \Bigg) + e^I_t, \nonumber
\end{align}
where the residuals follow the normal distributions $e_{t}^{\rho} \sim \mathcal{N}(0,(v^{\rho}_{p})^{2})$ and $e_{t}^{I} \sim \mathcal{N}(0,(v^{I}_{p})^{2})$.

A set of historical data of erythroblast growth from four different donors was analysed by \cite{glen2018mechanistic}  to provide the insight into autologous variation. Their investigation shows that only the growth rate has significant variability. 
In order to accommodate the {batch-to-batch variation}, 
we model the growth rate as normally distributed random variable $r^g_p\sim \mathcal{N}\left(\mu_{p}^g,(\sigma^g_{p})^2\right)$ with $p=1,2$. 
Thus, the collection of hybrid model parameters is $\pmb{\theta}=\{\mu^g_p, \sigma^g_p, k^s_p, k^c_p,r^d_p, v^{\rho}_p, v^I_p; \mbox{ with } p=1,2\}$ specifying the dynamics and variations in the growth and stationary phases. 

\end{sloppypar}

\subsection{Simulation Data Generation}
\label{subsec: simulation data}

Based on the ICH Q11 regulatory guidance on drug substance manufacture and quality \citep{mcdonald2012q11},
identifying and modeling the potential sources of
process variability is listed as a key aspect of QbD and process control. 
Thus, to assess the performance of proposed KG hybrid modeling and model-based RL framework, we create a simulation model capturing the key sources of uncertainty, including the variation in initial cell density, measurement error, batch-to-batch variation, and bioprocess noise. It works as a ground-truth simulator to generate ``real data" and assess the performance of proposed hybrid-RL framework. 

\begin{sloppypar}
The ground-truth cell culture process is modeled as stochastic differential equations,
\begin{align}
    \d\rho &= r^g_p \rho \Bigg (1 - \Big(1+e^{(k_p^s(k_p^c-I))} \Big) ^{-1} \Bigg )\d t+\sigma_n\mbox{d}W \label{equ:ODEtwophaseCell} \\
    \d I &= {\d\rho} - r^d_p I\d t+ \sigma_n\mbox{d}W \label{equ:ODEtwophaseI}
\end{align}
with random initial values (boundary conditions)
\begin{equation*}
        \rho_1 \sim \mathcal{N}(\mu_{\rho}, \sigma_{\rho}^2)\quad \text{ and } \quad I_1 = 0
\end{equation*}
\noindent where $r^g_p\sim\mathcal{N}\left(\mu_{p}^g,(\sigma^g_{p})^2\right)$ with $p=1,2$. 
In the computer simulation, the above stochastic differential equations (SDE) can be computed by discretizing the Wiener process with a timestep $\d t$ as $\d W \approx \sqrt{\d t}\mathcal{N}(0,1)$ \citep{bayram2018numerical,malham2010introduction}. 
We add measurement error to the measured cell density by $\rho_t\leftarrow\rho_t + e_m$ with the measurement error following a normal distribution $e_m\sim  \mathcal{N}(0,\sigma_m^2)$.

The variance in initial cell density could come from inadequate mixing, inoculum size, processing and storage times \citep{rain2021cryopreservation}. The bioprocess noise term $\sigma_n\d W$ represents the microbial cell-to-cell phenotypic diversity \citep{vasdekis2015origins}. 
We further incorporate batch-to-batch variation by considering the random effect on the growth rate, 
i.e., $r^g_p = \mu^g_{p}+e_p^g$, where $\mu_{p}^g$ represents the mean of growth rate and the random effect $e_{p}^g$ is modeled as 
a normal distribution with mean zero and standard deviation $\sigma_p^g$. 
\end{sloppypar}

\subsection{Process Model Inference and Validation}
\label{subsec:casestudyInference}


\begin{sloppy}
In the case study, we set the initial state of cell density $\rho_1 \sim \mathcal{N} (3,0.03^2)$ and the measurement uncertainty $\sigma_m = 0.2$.  By using the synthetic data generated from the ``ground-truth" simulation model (\ref{equ:ODEtwophaseCell})-- (\ref{equ:ODEtwophaseI}) in Section~\ref{subsec: simulation data}, we compare the prediction performance of the proposed Bayesian KG hybrid model with the deterministic mechanistic model estimated by using least square method (LS) under the situations with different levels of bioprocess inherent stochasticity and model uncertainty (i.e., different sizes of process observations $m = 3, 6, 20$ batches). For stochastic uncertainty, we are interested in the effect of batch-to-batch variation and bioprocess noise. We set the high and low batch-to-batch variation $\sigma_{p}^g = 0.016,0.008$ 
and set high and low bioprocess noise $\sigma_n = 0.03, 0.01$ for growth and stationary phases. 
We set the expected cell growth rate 
as $\mu^g_p= 0.057,0.0285$ for $p=1,2$. 
Other bioprocess parameters are set as $\{k^s, k^c,r^d\} = \{3.4,2.6,0.005\}$ for both phases.

We use the ABC-SMC sampling procedure in Algorithm~\ref{Algr:SMC-ABC}
to generate posterior samples of bioprocess model parameters. Let the number of particles $N = 200$, the ratio $\alpha = 0.5$, the number of replications $L = 20$, 
and the minimal accept rate $P_{acc_{min}} = 0.05$. 
The prior distributions of KG hybrid model parameters are set as: $\mu^g_p \sim U(0,0.2)$, $\sigma^g_p \sim U(0,0.05)$, $k_p^s \sim U(0,5)$, $k_p^c \sim U(0,5)$, $r_p^d \sim U(0,0.05)$, $v^\rho_p \sim U(0,0.2)$ and $v^I_p \sim U(0,0.2)$ for
$p=1,2$.


We conduct simulation experiments to assess the long-term prediction performance of the Bayesian KG and the ODE-based mechanistic model with the LS estimates. 
The $h$ time step ahead prediction error is defined as $$E_s(h,r)=\frac{1}{N}\sum_{i=1}^N\mbox{Err}_s(h, f_r)$$ where 
$\mbox{Err}_{s}(h,f_r)=|s_{h+1}^{(i)}-\widehat{s}_{1:h+1}^{(i)}|$ for $s \in \{\rho,I\}$ and $f_r\in \text{\{``hybrid'',``LS''\}}$ represents the estimated Bayesian KG hybrid model and LS mechanistic model fitted by ``real-world" historical trajectories with the size $m=3,6,20$ obtained in the $r$-th macro-replication. 
The current time step is $t=1$ and $\widehat{s}_{1:h+1}^{(i)}$ represents the $h$ step forward prediction.
For Bayesian KG, we use the mean of the predictive distribution of the predictor in eq.~(\ref{eq.prediction}) as the point predictor $\widehat{\rho}_{h+1}$ and $\widehat{I}_{h+1}$ for cell density $\rho_{h+1}$ and inhibitor concentration $I_{h+1}$ at $(h+1)$-th time step respectively.
We set the number of macro replications as 30 and test sample size $N=1000$.

Tables~\ref{table: prederror_rho} and ~\ref{table: prederror_I} record the mean absolute prediction error (MAE) with 95\% confidence interval (CI) obtained from $r=1,2,\ldots, 30$ macro-replications. We consider 3, 18 and 30 hours' look-ahead prediction (corresponding to $h = 1,6,10$), where the mean is defined as $E_s(h)=\frac{1}{30}\sum_{r=1}^{30}E_s(h,r)$, standard error is $\mbox{SE}_s(h)=\frac{1}{\sqrt{30}}\sqrt{\frac{1}{29}\sum_{r=1}^{30}(E_s(h,r)-E_s(h))^2}$.
The results demonstrate that our method can provide significantly better prediction accuracy under the situations with different levels of bioprocess noise (i.e., $\sigma_n=0.01,0.03$) and different sample sizes of ``real-world" observations (i.e., $m=3,6,20$). 
\end{sloppy}

\begin{table*}[ht]
\caption{The prediction error (MAE) of expected erythroblast cell density of hybrid model and LS.}
\label{table: prederror_rho}
\begin{tabular}{@{}cc|c|ccc|ccc@{}}
\toprule
\multicolumn{2}{c|}{Noise Level}         & {\multirow{2}{*}{\begin{tabular}[c]{@{}c@{}} $h$ \\ (hrs)\end{tabular}}}  & \multicolumn{3}{c|}{Hybrid}                                                                                                                                               & \multicolumn{3}{c}{LS}                                                                                                                                                  \\ \cmidrule{1-2} \cmidrule{4-9} 
b2b variation & process noise   &  & $m=3$                                                     & $m=6$                                                     & $m=20$                                                    & $m=3$                                                     & $m=6$                                                     & $m=20$                                                    \\ \midrule
high                     & $\sigma_n=0.01$ & \begin{tabular}[c]{@{}c@{}}$3$\\ $18$ \\ $30$\end{tabular} & \begin{tabular}[c]{@{}c@{}}0.12 $\pm$ 0.05\\ 0.60 $\pm$ 0.17 \\ 0.59 $\pm$ 0.16\end{tabular} & \begin{tabular}[c]{@{}c@{}}0.09 $\pm$ 0.03\\ 0.48 $\pm$ 0.10 \\ 0.40 $\pm$ 0.11\end{tabular} & \begin{tabular}[c]{@{}c@{}}0.06 $\pm$ 0.02\\ 0.26 $\pm$ 0.07 \\ 0.22 $\pm$ 0.06\end{tabular} & 
\begin{tabular}[c]{@{}c@{}}0.41 $\pm$ 0.19\\ 0.74 $\pm$ 0.15 \\ 0.65 $\pm$ 0.24\end{tabular} & \begin{tabular}[c]{@{}c@{}}0.59 $\pm$ 0.30\\ 0.57 $\pm$ 0.25 \\ 0.70 $\pm$ 0.36\end{tabular} & \begin{tabular}[c]{@{}c@{}}0.44 $\pm$ 0.22\\ 0.49 $\pm$ 0.23 \\ 0.84 $\pm$ 0.65\end{tabular} \\ \midrule
high                     & $\sigma_n=0.03$& \begin{tabular}[c]{@{}c@{}}$3$\\ $18$ \\ $30$\end{tabular} & \begin{tabular}[c]{@{}c@{}}0.21 $\pm$ 0.05\\ 1.07 $\pm$ 0.22 \\ 1.11 $\pm$ 0.24\end{tabular} & \begin{tabular}[c]{@{}c@{}}0.14 $\pm$ 0.03\\ 0.82 $\pm$ 0.15 \\ 0.74 $\pm$ 0.16\end{tabular} & \begin{tabular}[c]{@{}c@{}}0.08 $\pm$ 0.03\\ 0.48 $\pm$ 0.12 \\ 0.44 $\pm$ 0.12\end{tabular} & 
\begin{tabular}[c]{@{}c@{}}0.37 $\pm$ 0.20\\ 1.11 $\pm$ 0.28 \\ 1.57 $\pm$ 0.76\end{tabular} & \begin{tabular}[c]{@{}c@{}}0.40 $\pm$ 0.19\\ 0.93 $\pm$ 0.31 \\ 1.09 $\pm$ 0.41\end{tabular} & \begin{tabular}[c]{@{}c@{}}0.36 $\pm$ 0.23\\ 0.83 $\pm$ 0.34 \\ 0.93 $\pm$ 0.45\end{tabular} \\ \midrule
low                      & $\sigma_n=0.01$ & \begin{tabular}[c]{@{}c@{}}$3$\\ $18$ \\ $30$\end{tabular} & \begin{tabular}[c]{@{}c@{}}0.10 $\pm$ 0.03\\ 0.48 $\pm$ 0.12 \\ 0.47 $\pm$ 0.11\end{tabular} & \begin{tabular}[c]{@{}c@{}}0.07 $\pm$ 0.02\\ 0.38 $\pm$ 0.09 \\ 0.30 $\pm$ 0.08\end{tabular} & \begin{tabular}[c]{@{}c@{}}0.04 $\pm$ 0.01\\ 0.27 $\pm$ 0.06 \\ 0.16 $\pm$ 0.04\end{tabular} & 
\begin{tabular}[c]{@{}c@{}}0.38 $\pm$ 0.23\\ 0.43 $\pm$ 0.13 \\ 0.45 $\pm$ 0.11\end{tabular} & \begin{tabular}[c]{@{}c@{}}0.54 $\pm$ 0.26\\ 0.35 $\pm$ 0.12 \\ 0.52 $\pm$ 0.23\end{tabular} & \begin{tabular}[c]{@{}c@{}}0.38 $\pm$ 0.20\\ 0.28 $\pm$ 0.11 \\ 0.32 $\pm$ 0.15\end{tabular} \\ \midrule
low                      & $\sigma_n=0.03$ & \begin{tabular}[c]{@{}c@{}}$3$\\ $18$ \\ $30$\end{tabular} & \begin{tabular}[c]{@{}c@{}}0.18 $\pm$ 0.06\\ 1.00 $\pm$ 0.20 \\ 1.04 $\pm$ 0.28\end{tabular} & \begin{tabular}[c]{@{}c@{}}0.13 $\pm$ 0.03\\ 0.69 $\pm$ 0.15 \\ 0.65 $\pm$ 0.17\end{tabular} & \begin{tabular}[c]{@{}c@{}}0.04 $\pm$ 0.01\\ 0.27 $\pm$ 0.06 \\ 0.16 $\pm$ 0.04\end{tabular} & 
\begin{tabular}[c]{@{}c@{}}0.68 $\pm$ 0.34\\ 1.27 $\pm$ 0.36 \\ 1.40 $\pm$ 0.48\end{tabular} & \begin{tabular}[c]{@{}c@{}}0.46 $\pm$ 0.23\\ 1.47 $\pm$ 1.41 \\ 1.61 $\pm$ 1.40\end{tabular} & \begin{tabular}[c]{@{}c@{}}0.31 $\pm$ 0.18\\ 0.55 $\pm$ 0.16 \\ 0.66 $\pm$ 0.20\end{tabular} \\ \bottomrule
\end{tabular}
\end{table*}

\begin{table*}[ht]
\caption{The prediction error (MAE) of expected inhibitor accumulation fo hybrid model and LS.}
\label{table: prederror_I}
\begin{tabular}{@{}cc|c|ccc|ccc@{}}
\toprule
\multicolumn{2}{c|}{Noise Level}         & {\multirow{2}{*}{\begin{tabular}[c]{@{}c@{}}Time \\ (hrs)\end{tabular}}}  & \multicolumn{3}{c|}{Hybrid}                                                                                                                                               & \multicolumn{3}{c}{LS}                                                                                                                                                  \\ \cmidrule{1-2} \cmidrule{4-9} 
b2b variation & process noise   &  & $m=3$                                                     & $m=6$                                                     & $m=20$                                                    & $m=3$                                                     & $m=6$                                                     & $m=20$                                                    \\ \midrule
high                     & $\sigma_n=0.01$ & \begin{tabular}[c]{@{}c@{}}$3$\\ $18$ \\ $30$\end{tabular} & \begin{tabular}[c]{@{}c@{}}0.13 $\pm$ 0.05\\ 0.73 $\pm$ 0.19 \\ 1.23 $\pm$ 0.17\end{tabular} & \begin{tabular}[c]{@{}c@{}}0.10 $\pm$ 0.03\\ 0.49 $\pm$ 0.15 \\ 1.16 $\pm$ 0.12\end{tabular} & \begin{tabular}[c]{@{}c@{}}0.06 $\pm$ 0.02\\ 0.51 $\pm$ 0.11 \\ 1.12 $\pm$ 0.09\end{tabular} & 
\begin{tabular}[c]{@{}c@{}}0.41 $\pm$ 0.20\\ 1.82 $\pm$ 0.32 \\ 2.55 $\pm$ 0.33\end{tabular} & \begin{tabular}[c]{@{}c@{}}0.59 $\pm$ 0.30\\ 2.06 $\pm$ 0.26 \\ 2.53 $\pm$ 0.35\end{tabular} & \begin{tabular}[c]{@{}c@{}}0.44 $\pm$ 0.22\\ 2.09 $\pm$ 0.30 \\ 2.60 $\pm$ 0.41\end{tabular} \\ \midrule
high                     & $\sigma_n=0.03$& \begin{tabular}[c]{@{}c@{}}$3$\\ $18$ \\ $30$\end{tabular} & \begin{tabular}[c]{@{}c@{}}0.20 $\pm$ 0.06\\ 1.12 $\pm$ 0.25 \\ 1.33 $\pm$ 0.25\end{tabular} & \begin{tabular}[c]{@{}c@{}}0.15 $\pm$ 0.04\\ 0.88 $\pm$ 0.19 \\ 1.21 $\pm$ 0.20\end{tabular} & \begin{tabular}[c]{@{}c@{}}0.08 $\pm$ 0.03\\ 0.74 $\pm$ 0.16 \\ 1.22 $\pm$ 0.11\end{tabular} & 
\begin{tabular}[c]{@{}c@{}}0.37 $\pm$ 0.20\\ 1.78 $\pm$ 0.38 \\ 2.54 $\pm$ 0.41\end{tabular} & \begin{tabular}[c]{@{}c@{}}0.40 $\pm$ 0.19\\ 2.08 $\pm$ 0.27 \\ 2.79 $\pm$ 0.30\end{tabular} & \begin{tabular}[c]{@{}c@{}}0.36 $\pm$ 0.24\\ 2.23 $\pm$ 0.30 \\ 2.74 $\pm$ 0.35\end{tabular} \\ \midrule
low                      & $\sigma_n=0.01$ & \begin{tabular}[c]{@{}c@{}}$3$\\ $18$ \\ $30$\end{tabular} & \begin{tabular}[c]{@{}c@{}}0.10 $\pm$ 0.03\\ 0.52 $\pm$ 0.12 \\ 1.03 $\pm$ 0.14\end{tabular} & \begin{tabular}[c]{@{}c@{}}0.07 $\pm$ 0.02\\ 0.36 $\pm$ 0.10 \\ 1.07 $\pm$ 0.09\end{tabular} & \begin{tabular}[c]{@{}c@{}}0.04 $\pm$ 0.01\\ 0.34 $\pm$ 0.07 \\ 1.02 $\pm$ 0.08\end{tabular} & 
\begin{tabular}[c]{@{}c@{}}0.38 $\pm$ 0.23\\ 2.03 $\pm$ 0.27 \\ 2.90 $\pm$ 0.24\end{tabular} & \begin{tabular}[c]{@{}c@{}}0.54 $\pm$ 0.26\\ 2.84 $\pm$ 0.22 \\ 2.91 $\pm$ 0.23\end{tabular} & \begin{tabular}[c]{@{}c@{}}0.39 $\pm$ 0.20\\ 2.30 $\pm$ 0.22 \\ 2.95 $\pm$ 0.22\end{tabular} \\ \midrule
low                      & $\sigma_n=0.03$ & \begin{tabular}[c]{@{}c@{}}$3$\\ $18$ \\ $30$\end{tabular} & \begin{tabular}[c]{@{}c@{}}0.18 $\pm$ 0.05\\ 1.04 $\pm$ 0.22 \\ 1.27 $\pm$ 0.24\end{tabular} & \begin{tabular}[c]{@{}c@{}}0.13 $\pm$ 0.04\\ 0.75 $\pm$ 0.17 \\ 1.15 $\pm$ 0.18\end{tabular} & \begin{tabular}[c]{@{}c@{}}0.07 $\pm$ 0.02\\ 0.62 $\pm$ 0.13 \\ 1.07 $\pm$ 0.11\end{tabular} & 
\begin{tabular}[c]{@{}c@{}}0.68 $\pm$ 0.34\\ 2.17 $\pm$ 0.30 \\ 2.63 $\pm$ 0.36\end{tabular} & \begin{tabular}[c]{@{}c@{}}0.46 $\pm$ 0.23\\ 2.15 $\pm$ 0.29 \\ 2.67 $\pm$ 0.30\end{tabular} & \begin{tabular}[c]{@{}c@{}}0.31 $\pm$ 0.18\\ 2.13 $\pm$ 0.28 \\ 2.68 $\pm$ 0.33\end{tabular} \\ \bottomrule
\end{tabular}
\end{table*}


\subsection{Medium Full Exchange Decision Making}
\label{subsec:medium exchange}

In this section, we focus on finding the optimal time to fully exchange the medium with fresh medium. 
Execution of this action (exchanging medium) will immediately reset the concentration of inhibitor $I$ to its initial value (i.e., 0 in fresh medium).
Let $I_{t} = a_t I_t$ represent the post-exchanged concentration of inhibitor. The medium replacement decision $a_t$ is a binary variable: $a_t = 0$ denoting the full exchange of medium at step $t$; $a_t=1$, otherwise.

\begin{figure*}[pos=ht,width=1\textwidth,align=\centering]
    \begin{center}
    \includegraphics[width=1\textwidth]{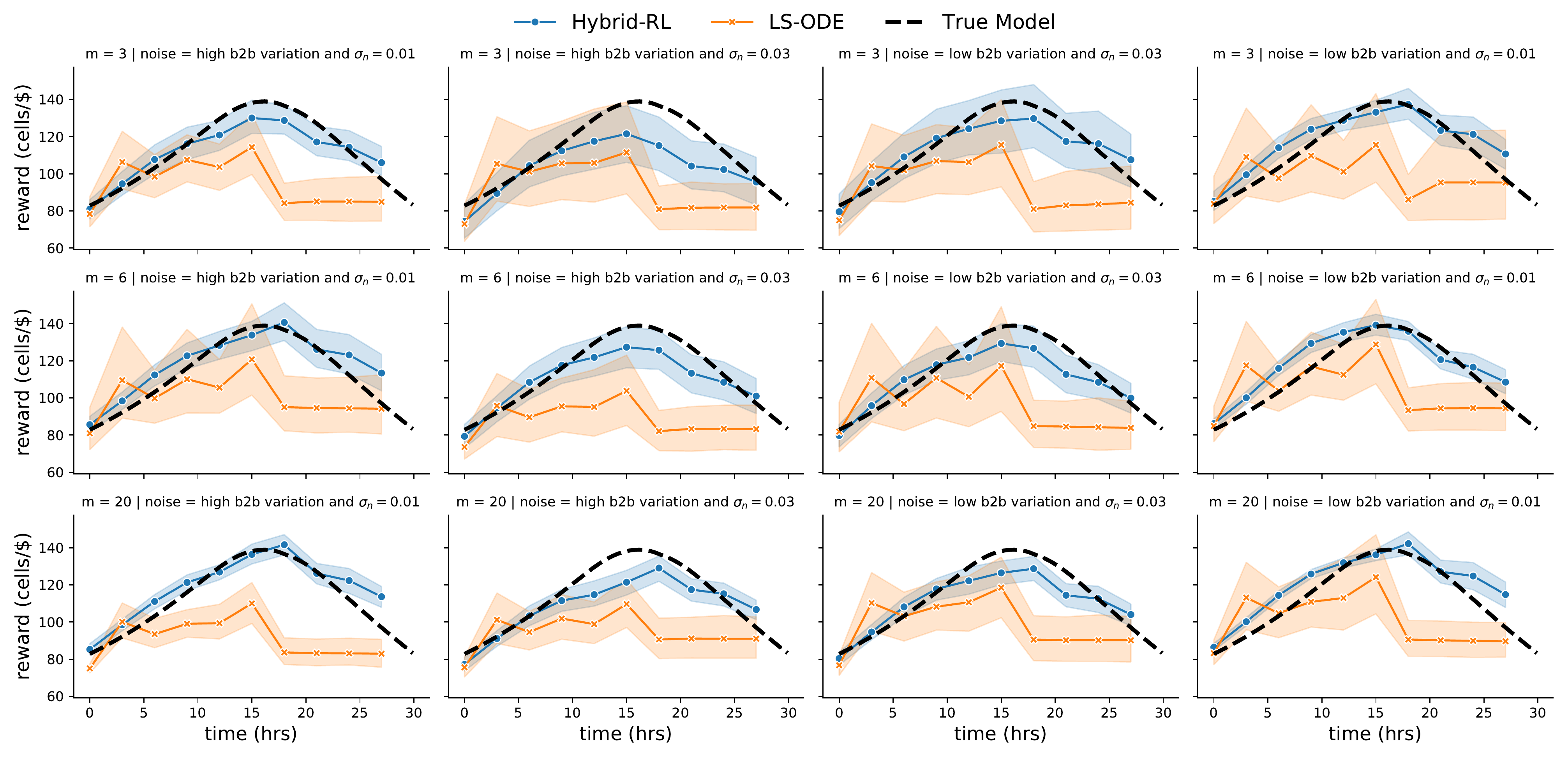}
    \caption{Performance comparison of hybrid model based reinforcement learning (hybrid-RL) and ODE mechanism model fitted by least square method (LS-ODE) in 30 macro-replications. The optimized reward with respect to the time of medium exchange are illustrated for hybrid-RL (point marker)), LS-ODE (``x'' marker) and ``ground-truth'' model (solid line). The validated models are used to optimize the media exchange time for cells to be produced with optimal cost efficiency given specified costs of cell culture medium and operational facility time at a given production scale (100L). The number of cells produced per dollar for a given time point of media exchange are calculated for any operating cost and consumable cost (here \$150/hr operating time cost and \$10/L of consumable cost per medium exchange).}\label{fig:medium exchange}
      \end{center}
      \vspace{-0.3in}
\end{figure*}

The total operational cost, denoted by $C$, includes the costs related to facility time and media \citep{glen2018mechanistic},
\begin{equation}
    C(T,M) = C_t T + C_m M \label{equ:opercost},
\end{equation}
where $C_t$ is the per unit facility use time-cost, $C_m$ is the per unit media cost, $T$ is the total cell culture time (in hours), and $M$ is the volume of the medium (L). As the time step $\Delta t=3$ hours, to avoid confusion, we use $H$ to represent the last time step or index of cell culture process and use $T$ to represent the total cell culture time (in hour).

The reward function is defined by cell yield per cost --- the efficiency of the system during the $T$ hours ($H$ time steps) cell culture \citep{glen2018mechanistic},
\begin{align}
&r_t = 0 \quad \text{with} \quad 0\leq t\leq H \nonumber\\
&    r_{H+1}(\pmb s_{H+1}, a_{H+1}=\text{``Harvest''}) 
= \frac{M(\rho_T-\rho_0)}{C(T,M)} \nonumber
\end{align}
where $\rho_T$ represents the cell density at the $T$-th hour. As it is certainly suboptimal to exchange medium at harvest time (30-th hour), we only consider step $t\leq 10$ since $\Delta t=3$ hours. Then the optimal timing of media exchange can be obtained by Algorithm~\ref{Algr: optimization}. We set the harvest time as 30 hours ($T=30$ corresponding to time step $t=H+1=11$) 
and the medium is fully exchanged up to one time in decision hours $\{0,3,\ldots,27\}$. 


\begin{table*}[!th]
\caption{Medium exchange cost efficiency (cells/\$) of hybrid model based reinforcement learning and ODE mechanism model based brute-force screening approach. Cost efficiency are summarized as the mean (standard error) of cells per dollar across 30 macro-replications.}\label{table: medium exchange}
\begin{tabular}{@{}cc|ccc|ccc@{}}
\toprule
\multicolumn{2}{c|}{Noise Level}           & \multicolumn{3}{c|}{Hybrid-RL}                                                                                                                                               & \multicolumn{3}{c}{LS-ODE}                                                                                                                                                  \\ \midrule
batch-to-batch variation & process noise   & $m=3$                                                     & $m=6$                                                     & $m=20$                                                    & $m=3$                                                     & $m=6$                                                     & $m=20$                                                    \\ \midrule
high                     & $\sigma_n=0.01$ & \begin{tabular}[c]{@{}c@{}}115.07\\ (1.48)\end{tabular} & \begin{tabular}[c]{@{}c@{}}117.99\\ (1.46)\end{tabular} & \begin{tabular}[c]{@{}c@{}}121.64\\ (1.44)\end{tabular} & \begin{tabular}[c]{@{}c@{}}111.11\\ (2.18)\end{tabular} & \begin{tabular}[c]{@{}c@{}}108.67\\ (2.14)\end{tabular} & \begin{tabular}[c]{@{}c@{}}112.67\\ (2.28)\end{tabular} \\ \midrule
high                     & $\sigma_n=0.03$ & \begin{tabular}[c]{@{}c@{}}113.74\\ (1.58)\end{tabular} & \begin{tabular}[c]{@{}c@{}}114.81\\ (1.47)\end{tabular} & \begin{tabular}[c]{@{}c@{}}114.58\\ (1.49)\end{tabular} & \begin{tabular}[c]{@{}c@{}}111.11\\ (2.18)\end{tabular} & \begin{tabular}[c]{@{}c@{}}114.40\\ (2.47)\end{tabular} & \begin{tabular}[c]{@{}c@{}}108.85\\ (1.69)\end{tabular} \\ \midrule
low                      & $\sigma_n=0.01$ & \begin{tabular}[c]{@{}c@{}}115.74\\ (1.39)\end{tabular} & \begin{tabular}[c]{@{}c@{}}119.16\\ (1.42)\end{tabular} & \begin{tabular}[c]{@{}c@{}}122.17\\ (1.38)\end{tabular} & \begin{tabular}[c]{@{}c@{}}110.12\\ (1.75)\end{tabular} & \begin{tabular}[c]{@{}c@{}}111.52\\ (2.33)\end{tabular} & \begin{tabular}[c]{@{}c@{}}112.70\\ (2.27)\end{tabular} \\ \midrule
low                      & $\sigma_n=0.03$ & \begin{tabular}[c]{@{}c@{}}114.98\\ (1.54)\end{tabular} & \begin{tabular}[c]{@{}c@{}}115.19\\ (1.39)\end{tabular} & \begin{tabular}[c]{@{}c@{}}116.00\\ (1.36)\end{tabular} & \begin{tabular}[c]{@{}c@{}}111.25\\ (2.02)\end{tabular} & \begin{tabular}[c]{@{}c@{}}111.87\\ (2.26)\end{tabular} & \begin{tabular}[c]{@{}c@{}}110.66\\ (2.06)\end{tabular} \\ \bottomrule
\end{tabular}
\end{table*}

\begin{sloppypar}
Following the study in \cite{glen2018mechanistic}, we consider an exchange time optimization problem where cells per cost is calculated for running a 100L bioreactor from the starting density $3\times 10^6$/mL with hypothetical facility and equipment operating costs of \$150/hr and cell culture medium costs of \$10/L. We solve this problem by using the proposed KG hybrid model based Bayesian reinforcement learning (hybrid-RL) and the ODE-based mechanistic model fitted by least square method (LS-ODE). We compare their performance in the situations with different levels of batch-to-batch (b2b) variation and bioprocess noise, as well as different sample sizes (i.e., $m=3,6,20$) of the ``ground-truth'' cell culture process observations. Similar to \cite{glen2018mechanistic}, the optimization of LS-ODE was performed via an exhaustive search of decision space. 
\end{sloppypar}

\begin{sloppypar}

We illustrate the cost efficiency of full medium exchange at each decision time across different levels of uncertainty and sample size in Figure~\ref{fig:medium exchange}. The markers represent the rewards/cost efficiencies at the corresponding decision time and the translucent bands around marks are the 95\% confidence intervals estimated by bootstrapping. 
The results are obtained based on 30 macro-replications.
The cost efficiency curves estimated by LS-ODE have large discrepancies to those from ``ground-truth'' model and they have significantly wider confidence bands, indicating that this model suffers from the high variation between macro-replications. In addition, we observe that hybrid-RL has a shrinking trend of the confidence intervals as the sample size $m$ increases. 
\end{sloppypar}

\begin{sloppypar}
We summarize the performance of optimal decisions obtained from hybrid-RL and LS-ODE approaches in Table~\ref{table: medium exchange}. Overall, hybrid-RL  outperforms LS-ODE (with higher reward/cost efficiency) in all different levels of variations and sample sizes. 
It shows a strong evidence that hybrid-RL is more robust than the regular deterministic mechanistic model fitted by least-square method. 
Another interesting observation is that the optimal medium exchange decisions obtained by LS-ODE are not as bad as the estimated cost efficiency curve in Figure~\ref{fig:medium exchange}. It is because the response curve is relative flat (range from 104 cells/\$ to 167 cells/\$). 
To further evaluate the performance, we consider another optimization problem, ``cell culture expansion'' with the same simulated data and process models used in Section~\ref{subsec:culture expansion}.
\end{sloppypar}

\begin{table*}[ht]
\caption{Culture expansion profit (\$) of hybrid-RL and ODE mechanism model based brute-force screening approach. Profit are summarized as the mean (standard error) of dollar across 30 macro-replications.}\label{table: culture expansion}
\begin{tabular}{@{}cc|ccc|ccc@{}}
\toprule
\multicolumn{2}{c|}{Noise Level}           & \multicolumn{3}{c|}{Hybrid-RL}                                                                                                                                               & \multicolumn{3}{c}{LS-ODE}                                                                                                                                                  \\ \midrule
batch-to-batch variation & process noise   & $m=3$                                                     & $m=6$                                                     & $m=20$                                                    & $m=3$                                                     & $m=6$                                                     & $m=20$                                                    \\ \midrule
high                     & $\sigma_n=0.01$ & \begin{tabular}[c]{@{}c@{}}7317.24\\ (352.37)\end{tabular} & \begin{tabular}[c]{@{}c@{}}7588.15\\ (322.53)\end{tabular} & \begin{tabular}[c]{@{}c@{}}7892.84\\ (69.65)\end{tabular} & \begin{tabular}[c]{@{}c@{}}5677.62 \\ (389.24)\end{tabular} & \begin{tabular}[c]{@{}c@{}}5944.25\\ (403.52)\end{tabular} & \begin{tabular}[c]{@{}c@{}}6030.83\\ (399.61)\end{tabular} \\ \midrule
high                     & $\sigma_n=0.03$ & \begin{tabular}[c]{@{}c@{}}6888.86\\ (693.07)\end{tabular} & \begin{tabular}[c]{@{}c@{}}7266.23\\ (393.66)\end{tabular} & \begin{tabular}[c]{@{}c@{}}7689.45\\ (143.11)\end{tabular} & \begin{tabular}[c]{@{}c@{}}-2259.01\\ (704.82)\end{tabular} & \begin{tabular}[c]{@{}c@{}}1026.60\\ (484.50)\end{tabular} & \begin{tabular}[c]{@{}c@{}}2454.22\\ (262.15)\end{tabular} \\ \midrule
low                      & $\sigma_n=0.01$ & \begin{tabular}[c]{@{}c@{}}7800.60\\ (151.01)\end{tabular} & \begin{tabular}[c]{@{}c@{}}7955.38\\ (75.78)\end{tabular} & \begin{tabular}[c]{@{}c@{}}8035.71\\ (52.48)\end{tabular} & \begin{tabular}[c]{@{}c@{}}6115.31\\ (413.23)\end{tabular} & \begin{tabular}[c]{@{}c@{}}6193.70 \\ (381.83)\end{tabular} & \begin{tabular}[c]{@{}c@{}}6417.39\\ (389.06)\end{tabular} \\ \midrule
low                      & $\sigma_n=0.03$ & \begin{tabular}[c]{@{}c@{}}7414.34\\ (334.41)\end{tabular} & \begin{tabular}[c]{@{}c@{}}7572.99\\ (329.62)\end{tabular} & \begin{tabular}[c]{@{}c@{}}7974.35\\ (76.15)\end{tabular} & \begin{tabular}[c]{@{}c@{}}5978.52\\ (389.96)\end{tabular} & \begin{tabular}[c]{@{}c@{}}6126.60\\ (393.61)\end{tabular} & \begin{tabular}[c]{@{}c@{}}6225.02\\ (395.50)\end{tabular} \\ \bottomrule
\end{tabular}
\end{table*}


\subsection{Cell Culture Expansion Scheduling}
\label{subsec:culture expansion}

In this section, we focus on optimizing the number and timing of cell expansions. For each expansion, the original batch is scaled up to a $n$ times larger cell culture vessel filling with fresh medium. 
The cell density $\rho$ and the concentration of inhibitor $I$ decrease to $1/n$ of original batch immediately after each scale-up.  
If the culture expansion action is taken, that is $a_t=1$, then the post-exchanged cell density becomes $\rho_t = \rho_t/n$ and the level of post-exchanged inhibitor becomes $I_t = I_t/n$; otherwise, $a_t=0$. The optimal decision for the cell culture scale-up problem can be solved by using Algorithm~\ref{Algr: optimization}.
The expansion decision can be made every 3 hours starting from the third hour until the harvest time at the 30-th hour ($T=30$). 
Therefore, the maximal number of potential culture expansions is 9 if the expansion decisions are made at all candidate decision hours. 

In terms of the reward function, we use the same total operational cost as presented in Section~\ref{subsec:medium exchange}, and further introduce the revenue of cell therapy production based on unit price of cells denoted by $P_c$ as follows
\begin{equation*}
    K(\rho_T,\xi,n) = P_c\times \rho_T \times n^\xi
\end{equation*}
where $\xi$ is the total number of cell culture expansion and $\rho_T$ is the cell density at harvest hour $T$. Here we set expansion scale $n=4$ and set the price $P_c=\$2 \times 10^{-6}$  per cell. 

The reward function is then defined by the difference between revenue and cost as,
\begin{align}
&r_t = 0 \text{ with } 0\leq t \leq H \nonumber\\
    &r_{H+1}(\pmb{s}_{H+1}, {a}_{H+1}
     =\text{``Harvest''})\nonumber\\
     &\quad \quad =K(\rho_T, \xi,n) - C(T,M) \nonumber
\end{align}
where $C(T,M)$ is defined by \eqref{equ:opercost}. In this case, we expand the culture in a 1L seed bioreactor multiple times to achieve maximal profit. We set the initial cell density as $3 \times 10^6$/mL, hypothetical facility and equipment operating costs as \$150/hr and cell culture medium costs as \$10/L.

\begin{sloppypar}
We summarize the performance of hybrid-RL and LS-ODE methods in the situations with different variation levels and different sample sizes in Table~\ref{table: culture expansion}. It shows a strong evidence that hybrid-RL provides significantly better and more robust expansion decisions compared to LS-ODE. In all these settings, the mean profit of hybrid-RL obtained from 30 macro-replications are all greater than \$6000. Comparatively, the performance of LS-ODE method, with profits below \$6,500 in all cases, is significantly worse than hybrid model especially in the case with high batch-to-batch variation and bioprocess noise.
\end{sloppypar}

To better understand the optimality gap, we performed a side experiment to calculate the true optimal profit based on the ``ground-truth'' simulator and conduct the exhaustive search of all possible combinations of actions ($2^9$ different combinations) to find the optimal expansion decision. As a result, the true optimal action is found to be running one expansion at the 6-th hour (from 1 L to 4 L) with the corresponding maximal profit \$8593.85. By comparing the mean profit of hybrid-RL with the true expected optimal profit, it shows that the optimality gap of hybrid-RL is consistently lower than LS-ODE. Evidently, the largest optimality gap \$1704.99 occurs at the situation with high stochastic uncertainty (i.e., 
high batch-to-batch variation and bioprocess noise) and high model uncertainty (i.e., $m=3$). The smallest gap \$558.14 occurs at the situation with low inherent stochasticity and model uncertainty.

\section{Conclusions}
\label{sec:conclusion}

\begin{sloppypar}
The cell therapy manufacturing process stochastic uncertainty (i.e., batch-to-batch variation and bioprocess noise)
is often overlooked by the current mechanistic model literature. This can lead to sub-optimal and unreliable decision making. 
Thus, we introduce a probabilistic  knowledge graph (KG) hybrid model characterizing complex interactions of various sources of uncertainties and mechanism pathways connecting CPPs/CQAs.
It can leverage on the information from existing PDE/ODE-based mechanistic models and facilitate learning from heterogeneous data. 
This hybrid model can capture the important properties of integrated biomanufacturig processes, including nonlinear reactions, partially observed state, and time-varying dynamics.

To efficiently support process development and control, we create a KG hybrid model-based reinforcement learning accounting for both inherent stochasticity and model uncertainty. It can provide an insightful and reliable prediction on how the effects of decision inputs propagate through bioprocess mechanism pathways and impact on the output trajectory dynamics and variation.
Therefore, the proposed hybrid-RL framework can find optimal, robust, and interpretable decisions and control policies, which can overcome the key challenges of cell therapy manufacturing, i.e., high complexity, high uncertainty, and very limited process data. 
The empirical study demonstrates that the proposed framework can outperform the classical approach, especially under the situations with high inherent stochasticity and limited historical data which induces high model uncertainty.

\end{sloppypar}

\section*{Acknowledgements}
We acknowledge Richard D. Braatz (from Massachusetts Institute of Technology) and Cheng Li (from  National University of Singapore) for their insightful discussions and constructive feedback on this study. 

\bibliographystyle{cas-model2-names}
\bibliography{reference, proposal,proj_ref}

\end{document}